\documentclass[11pt,a4paper]{article}
\pdfoutput=1

\usepackage{jheppub}

\usepackage[dvipsnames]{xcolor}
\usepackage{amsmath,amsfonts,amssymb}
\usepackage{epsfig}
\usepackage{url}

\newcommand{\refeq}[1]{(\ref{#1})}
\newcommand{\nn}{\nonumber \\}
\newcommand{\beq}{\begin{equation}}
\newcommand{\eeq}{\end{equation}}
\newcommand{\bal}{\begin{align}}
\newcommand{\eal}{\end{align}}

\newcommand{\Lcal}{\mathcal{L}}
\newcommand{\Ocal}{\mathcal{O}}
\newcommand{\Dcal}{\mathcal{D}}

\newcommand{\eps}{\epsilon}
\renewcommand{\d}{\partial}
\newcommand{\hc}{\text{h.c.}}

\newcommand{\re}{\operatorname{Re}}
\newcommand{\res}{\operatorname{Res}}

\newcommand{\truncated}[2]{{[{#1}]}_{#2}}
\newcommand{\corrected}[2]{{\left<{#1}\right>}_{#2}}

\title{Field redefinitions in effective theories at higher orders}

\author[a]{J. C. Criado}
\author[a]{M. P\'erez-Victoria}

\affiliation[a]{
  CAFPE and Departamento de F\'{\i}sica Te\'orica y del Cosmos,\\
  Universidad de Granada, Campus de Fuentenueva, E-18071 Granada, Spain
}

\emailAdd{jccriadoalamo@ugr.es}
\emailAdd{mpv@ugr.es}

\abstract{
  The invariance of physical observables under redefinitions of the quantum
  fields is a well-known and important property of quantum field theory. We
  study perturbative field redefinitions in effective theories, paying special
  attention to higher-order effects and their impact on matching to an
  ultraviolet theory at the classical and quantum levels.
}

\begin{document}

\maketitle
\flushbottom

\section{Introduction}

Effective field theory provides a very general and precise framework to describe physical systems at the quantum level.
A particular effective quantum field theory can be defined {\it \`a la\/} Wilson by a quasi-local action,\footnote{``Quasi-local'' means that it is the integral of a Lagrangian density with a Fourier transform that is analytic in momenta (and fields) in some region containing the origin. So, it can be written in a derivative expansion as an infinite series of (local) operators. A ``local" action, in contrast, only has terms with a bounded number of derivatives.} a cutoff procedure and the value of the cutoff~\cite{Polchinski:1983gv}. Alternatively, it can be defined by a quasi-local action, a renormalization scheme and a renormalization scale~\cite{Weinberg:1995mt}. In this second approach, which we follow here, any intermediate regularization is removed and the quantum fluctuations explore arbitrary energy scales. 
The action that describes an effective theory is given in general by a linear combination of an infinite number of local operators, restricted to respect the required symmetries. It typically contains a characteristic scale $\Lambda$,  which controls the size of the different coefficients. At energies higher than this scale, either new physics or some new strong-coupling regime of the theory should show up, in such a way that unitarity is preserved. At any rate, what is most important from the low-energy point of view is that the predictions for a physical observable can be expressed as a power series in $E/\Lambda$, with $E$ indicating collectively the relevant energy-momentum scales in the process and other dimensionful scales appearing in the action, such as the masses of light particles. As long as $E \ll \Lambda$, the results can be approximated by keeping a finite number of terms in the $E/\Lambda$ expansion. Within this approximation, the theory can then be described by an action that is a polynomial in $1/\Lambda$. This effective action is local and depends only on a finite number of parameters. But, as we discuss in detail in the paper, it is not necessarily a simple truncation of the exact action. 

The description of a given quantum field theory in terms of an action and a renormalization scale (or a cutoff, in the Wilsonian approach) is highly redundant. Firstly, the renormalization group invariance represents a one-parameter redundancy: a change in the renormalization scale (or in the cutoff) can be compensated by a change in the action in such a way that the predictions of the theory are preserved.\footnote{More generally, any change of renormalization scheme can compensated by a change in the action.} Secondly, physical observables are invariant under redefinitions of the quantum fields.\footnote{Actually, the renormalization group invariance can be understood as the invariance under a particular type of  field redefinition~\cite{Latorre:2000qc}.} This property of quantum field theory is sometimes known as the equivalence theorem (not to be confused with the equivalence theorem in the Higgs mechanism). Different versions of this theorem, with different assumptions and in different contexts, have been proved and discussed in the literature~\cite{Chisholm:1961tha,Kamefuchi:1961sb,Divakaran:1963yxz,Kallosh:1972ap,Salam:1971sp,Ball:1993zy,Arzt:1993gz}. Here we have in mind the application of the effective field theory to the scattering of particles. In this context, the redundancy is given by the freedom in choosing interpolating fields that can create the relevant particles from the vacuum and be used to compute scattering amplitudes.

In the present work, we explore some aspects of local perturbative field redefinitions in effective field theories. By perturbative, we mean that the variation in the fields is treated as a perturbation. These redefinitions have the virtue of being automatically invertible with a local inverse, in a perturbative sense. Moreover, as shown, for instance, in~\cite{tHooft:1973wag} and reviewed below, their effect is particularly simple,  as the Jacobian of the transformation can be ignored in methods such as dimensional regularization. Most of the time, the change of the fields will be taken to be suppressed by some positive integer power of $1/\Lambda$. Then, treating it as a perturbation is actually implied by the perturbative expansion of the effective theory in powers of $1/\Lambda$. This kind of redefinition mixes different orders in the $1/\Lambda$ expansion of the effective action in a triangular fashion: the $n$-th order of the redefined action depends only on terms of order $m\leq n$ in the original one. 

Perturbative redefinitions are performed customarily by effective-theory practitioners in order to write general or particular effective actions, consistent with certain symmetries, in reduced forms~\cite{Georgi:1991ch, Arzt:1993gz,Grzadkowski:2003tf,Fox:2007in,AguilarSaavedra:2008zc,AguilarSaavedra:2009mx,Grzadkowski:2010es}. The idea is to eliminate part of the reparametrization redundancy by imposing a condition on the action. This is completely analogous to fixing a gauge in a gauge-invariant theory, and we will borrow this terminology.\footnote{In fact, this is more than a mere analogy: any quantum field theory has a BRST symmetry associated with field redefinitions~\cite{Alfaro:1989rx}.} A standard gauge-fixing condition is to require the vanishing of the coefficients of certain operators. As we review in section~\ref{sec:perturbative}, this can be achieved order by order in $1/\Lambda$ by perturbative redefinitions. If no linear combination of the remaining operators can be redefined away without violating the gauge-fixing condition, then these operators are said to form a non-redundant basis. 
When eliminating a certain term of order $n$, the change in the action at orders $m>n$ (and in the other terms at order $n$) can be absorbed into the operator coefficients of the original action, if it is completely general.\footnote{Note that the necessary redefinitions will always preserve the symmetries of the action, see appendix~\ref{app:covariance}.} From the purely effective point of view, there is often no need to track this redefinition of the coefficients, as they are free parameters to be determined experimentally. For this reason, among others, the ``higher-order effects'' of the field redefinition are usually ignored. Then, it turns out that the order-by-order algorithm to remove operators with perturbative redefinitions is equivalent to a very simple recipe: using the equations of motion of the action at lowest order $(n=0)$ in any of the terms of order $n\geq 1$~\cite{Georgi:1991ch}. 

However, in many situations it is crucial to know the dependence of the coefficients in the redefined action on the coefficients of the original one. This is the case, for instance, when one wants to translate the experimental limits on the coefficients in one basis of operators into limits on the coefficients of the operators in another (reduced) basis. Another common scenario is the one in which the coefficients in a certain effective action are known functions of the parameters of some ultraviolet (UV) completion of interest, and one wants to know the corresponding functional dependence of the operator coefficients in a particular non-redundant basis. In these situations, all the effects of the field redefinitions up to a certain order must be considered if the aimed precision requires a calculation to that order~\cite{Manohar:1997qy}.  Analyzing the perturbative structure of these effects---including the impact of quantum corrections and dimensionless couplings---is the main purpose of this work.  In particular, we clarify the relation between perturbative field redefinitions and the classical equations of motion, which still is, apparently, the source of some confusion. For example, it is well known that many of the corrections of order $n\geq 2$ generated by the perturbative redefinitions are missed by the recipe based on the lowest-order equations of motion.  One could try to improve this situation by including higher-order terms in the equation of motion, as done in~\cite{Jenkins:2017dyc,Barzinji:2018xvu}. We show, however, that the higher-order corrections induced by the redefinitions are not correctly recovered by this extended recipe. The essential reason is that the classical equations of motion only capture the first-order response of the action to variations of the fields. Therefore, using naively the equations of motion, with or without higher-order corrections, gives in general an action that is not equivalent to the original one at the second and higher orders. Imposing equations of motion is not the same as performing field redefinitions. 

Whether the higher-order terms, in particular those induced by field redefinitions, are significant or not in practice, depends on many factors: the experimental precision, the process to be calculated, the theory at hand, the value of $E/\Lambda$ and the value of the remaining parameters that appear in the action. For instance, it may happen that the first-order contributions are vanishing or very suppressed, due to some symmetry or to some dynamical reason. Then the second-order terms would give the leading correction~\cite{Arzt:1994gp,Hays:2018zze}. 

Taking into account the higher-order terms generated by field redefinitions is relevant, in particular, for the consistent perturbative matching of a local effective theory to a more fundamental UV theory with the same light degrees of freedom.  In this respect, we also study the impact on the effective theory of field redefinitions performed in the UV theory. We find that, non-trivially, the redefinitions of the light fields do not commute at the quantum level with the matching procedure. This is related to the contribution of heavy-light loops in the UV theory.

The article is organized as follows. In section~\ref{sec:reparametrization-invariance}, we review the effect of local redefinitions on quantum field theories for off-shell and on-shell quantities, paying special attention to the case of perturbative redefinitions.\footnote{Much of the content of this section can be found in Ref.~\cite{Arzt:1993gz}. We also clarify a couple of important details and summarize latter work on renormalization.} 
In section~\ref{sec:eom}, we discuss the relation between field redefinitions and the classical equations of motion. In section~\ref{sec:matching}, we examine how field redefinitions affect the matching of an effective theory to a more fundamental one. In section~\ref{sec:perturbative}, we analyze perturbative field redefinitions in which the perturbation is controlled by the same small parameters as the perturbative expansion of the effective theory. This refers mainly to the length scale $1/\Lambda$, but also to other dimensionless parameters that may enter in the effective theory, such as coupling constants or $1/4\pi$ factors related to loops. We also point out a few effects at higher orders in $1/\Lambda$ or in the loop expansion. In section~\ref{sec:example} we give a realistic example in the Standard Model effective field fheory (SMEFT) that illustrates some of the general results. In section~\ref{sec:conclusions} we extract the main conclusions of our study, including a proposal for the workflow in effective field theories. In appendices~\ref{app:free},~\ref{app:eom-counterexample} and ~\ref{app:simple-matching-counterexample} we present, respectively, a toy model illustrating the role of the Jacobian and the sources in the field transformation, a counterexample to the exact validity of eliminating operators proportional to the classical equations of motion and a sample calculation that proves the appearance of non-trivial effects of field redefinitions when a quasi-local action is truncated at a finite order in the $1/\Lambda$ expansion. Finally, appendix~\ref{app:covariance} shows that explicit gauge covariance is preserved by covariant  field redefinitions and is manifest in the exact equations of motion of a gauge theory. In particular, this implies that the corrections to the SMEFT equations of motion given in ref.~\cite{Barzinji:2018xvu} in terms of ordinary derivatives and gauge fields can be written in terms of field-strength tensors and covariant derivatives. 

In order to make it as didactical and self-contained as possible, our exposition follows a logical order and includes some results that have been given in diverse forms in the literature of the last forty years. In the same vein, we have provided simple examples that explicitly illustrate the less obvious effects of field redefinitions. To help readers locate the original material, we highlight next our main new results (including observations relevant for current phenomenological applications), together with the sections where they can be found:
\begin{enumerate}
\item We show that arbitrary local perturbative field transformations can be applied to fields with vanishing or non-vanishing vacuum expectation value. This is non-trivial, since the expectation values of the new fields are not simply given, at the quantum level, by the same transformation of the original expectation values. As an important corollary, covariant field redefinitions can be performed in theories with spontaneously broken symmetries before shifting the fields (which is actually common practice in the SMEFT). [Section~\ref{sec:reparametrization-invariance}]
\item We give a necessary condition (known to be sufficient) for a parameter in the action to be exactly redundant. [Section~\ref{sec:eom} (examples in section~\ref{sec:example} and appendix~\ref{app:free})]
\item We give a necessary condition (known to be sufficient) for a term in the action to be exactly redundant. [Section~\ref{sec:eom}]
\item We study how field redefinitions in a UV field theory are related to field redefinitions in the corresponding low-energy effective theory. We find that redefinitions of the light fields do not commute with matching at the quantum level. [Section~\ref{sec:matching} and appendix~\ref{app:simple-matching-counterexample}]
\item We analyze how perturbative field redefinitions affect power counting. [Section~\ref{sec:perturbative} (example in section~\ref{sec:example})]
\item We find which contributions are missed when the classical equations of motion are used to simplify the action, and show that these contributions cannot be neglected at higher orders. [Sections~\ref{sec:eom} and~\ref{sec:perturbative} (subsection~\ref{subsec:nonredundant}) (examples in section~\ref{sec:example} and appendix~\ref{app:eom-counterexample})]
\item We point out a few non-trivial consequences at higher orders of the usage of field redefinitions in effective theories, which are relevant for current phenomenological applications. [Section~\ref{sec:perturbative} (example in section~\ref{sec:example})]
\end{enumerate}

\section{Reparametrization invariance}
\label{sec:reparametrization-invariance}

Consider a quantum field theory described by a classical action $S$, which is a
local or quasi-local functional of a set of quantum fields $\phi^i$, represented
collectively by $\phi$. We follow the convention of indicating the adjoints of
complex fields with distinct labels $i$, in such a way that a sum over $i$
includes both a field and its adjoint, if not real. Furthermore, we use the
compact DeWitt notation $\phi^\alpha = \phi^i(x)$, with repeated collective indices
indicating also integration over the space-time variables. Let $Z[J]$ be the
generating function of multiple-point Green functions, with sources $J_i$ for
each field $\phi^i$. In terms of a Lagrangian path-integral,
\begin{equation}
  \label{eq:original-Z}
  Z[J]
  =
  \int \Dcal \phi
  \exp{
    \left(
      i S[\phi]
      + J_\alpha \phi^\alpha
    \right),
  }
\end{equation}
with the normalization $Z[0]=1$. A possible non-trivial determinant in the
measure (necessary for instance in a non-linear sigma model) is assumed to be
included in the action $S$. This is just a formal expression and a regularization or
renormalization procedure is necessary to give a precise meaning to it and to
the following manipulations. Now, let us perform a change of integration
variables $\phi \to \phi^\prime = F(\phi)$, where $F$ is an invertible
function. 
Ignoring regularization and renormalization for the moment, we get
\begin{equation}
  \label{eq:transformedZ}
  Z[J]
    =
  \int \Dcal \phi
  \det{
    \left( \frac{\delta F}{\delta \phi} \right)
  }
  \exp{
    \left(
      i S[F(\phi)]
      + J_\alpha F^\alpha(\phi)
    \right)
  }.
\end{equation}
So, the generating function is invariant under a field redefinition in the
action, $S^\prime[\phi]=S[F(\phi)]$, if the redefinition is accompanied by
the corresponding Jacobian factor and the corresponding change in the source
terms, as specified by eq.~\refeq{eq:transformedZ}. Usually, the transformation $F$ is taken to respect the symmetry and hermiticity properties of the original action, although this is not strictly necessary: as long as the transformation is invertible, the change of variables is valid and the generating function will remain invariant (see nonetheless comments in~\cite{Passarino:2016saj}).
At the regularized level, the same equation holds, possibly with a non-trivial
redefinition of the regularization.%
\footnote{ For instance, a regularization that modifies only the quadratic terms
  will modify the interactions after a non-linear field redefinition. The change in the regulator accounts for the additional terms found in ~\cite{Gervais:1976ws}. The
  definition of methods such as dimensional regularization, which preserve the
  formal relations between propagators and
  interactions~\cite{Breitenlohner:1977hr}, needs not be changed.%
}
In the following we consider only regularizations in which the regulator is
dimensionless, and therefore does not interfere with the cutoff expansion in
effective theories. Dimensional regularization belongs to this class. 

Both the Jacobian and the modified source terms are required for $Z$ to remain
invariant. In particular, they are necessary to cancel possible higher-order
poles, as illustrated in appendix~\ref{app:free}. Fortunately, they can
be neglected under certain circumstances, as we now
review. This is the usual statement of the equivalence theorem.

The Jacobian of the transformation can be written in terms of ghost fields $c$,
$\bar{c}$:
\begin{equation}
  \label{eq:jacobian-ghosts}
  \det{\frac{\delta F}{\delta \phi}}
  =
  \int \Dcal \bar{c} \Dcal c
  \exp{
    \left(
      -i \bar{c}_\alpha \frac{\delta F^\alpha}{\delta \phi^\beta} c^\beta
    \right)
  }.
\end{equation}
In the following we consider only local transformations, with $F^{ix}(\phi)$
depending analytically on the value of the fields and their derivatives, up to a
finite order, at the point $x$. Then the Jacobian in terms of ghosts can be
simply added to the action, which remains (quasi) local. In general, this contribution
to the action has a non-trivial effect (see appendix~\ref{app:free}). However,
most applications involve perturbative field redefinitions
\begin{equation}
  \label{eq:pfr}
  F(\phi) = \phi + \lambda G(\phi), 
\end{equation}
where $G$ is analytic in $\lambda$ and all terms proportional to
positive powers of $\lambda$ are to be treated as interactions in perturbation theory. Then, the inverse of the transformation is also local. Moreover,
the ghost propagator is equal to the identity and the ghost loops only contain
insertions of $\delta G(\phi)/\delta \phi^\alpha$, which by the locality assumption are polynomials of the
internal momenta. Therefore the ghost loops will integrate
to zero in dimensional regularization~\cite{tHooft:1973wag}. Then, in dimensional regularization (and in
any regularization with this property), the Jacobian of a local, perturbative
transformation is equal to the identity and the ghosts can be ignored. In the same manner, in the perturbative treatment no unphysical poles will appear in the propagators of the other (redefined) fields, and all the contributions that were cancelled by ghost loops will also vanish in dimensional regularization.  Note that it is crucial for consistency that all the quadratic terms in $S^\prime$ that vanish as
$\lambda\to 0$ be treated as interactions, i.e. not be resummed into the propagators of that theory. Check appendix~\ref{app:free} for an explicit example of this point.

The change in the source terms is important for the invariance of off-shell
quantities, but thanks to the LSZ reduction formula~\cite{Lehmann:1954rq} it has no impact on the S
matrix, at least for local perturbative redefinitions. To understand this, note
first that the poles of the momentum-space two-point function of any operator
$\Ocal$ are equal to the physical masses $m_a$ of the particles $a$ that this
operator can create from the vacuum. The probability amplitude of creating
particle $a$ with momentum $p$,
$ \sqrt{\mathcal{Z}^a_{\Ocal}}
:=
\left< a p \right| \Ocal(0) \left| 0 \right>
\neq
0
$,
is given by the residues at the poles. The operator $\Ocal$ is then a valid
interpolating field that can be used in the reduction formula to find S-matrix
elements involving any of the particles $a$, with wave-function renormalization
given by $\sqrt{\mathcal{Z}^a_{\Ocal}}$. For a perturbative field
redefinition~eq.~\refeq{eq:pfr},
\begin{equation}
  \left< a p \right| F^{i 0}(\phi) \left| 0
  \right> = \left< a p \right| \phi^{i0} \left| 0 \right> + O(\lambda).
\end{equation}
Therefore, if $\mathcal{Z}^a_{\phi^i} \neq 0$ when $\lambda \to 0$, then
$\mathcal{Z}^a_{F^i(\phi)} \neq 0$. Hence, $F^i(\phi)$ is
also a valid interpolating field for the particle $a$. Moreover, because the
physical masses of the particles do not know about the field representation, the
poles in the two-point function will remain the same at any order. In terms of
generating functionals, all this means that $Z$ and $Z^\prime$, with
\begin{equation}
  \label{eq:on-shell}
  Z^\prime[J]
  =
  \int \Dcal\phi
  \det{ \left( \frac{\delta F}{\delta \phi} \right)}
  \exp{
    \left( i S^\prime[\phi] + J_\alpha \phi^\alpha \right),
  }
\end{equation}
give rise to the same S matrix. We will say
that they are equivalent on-shell and write $Z \sim Z^\prime$.

Let us emphasize that this result holds for a general perturbative
redefinition~\cite{Arzt:1993gz}. The function $G$ in eq.~\refeq{eq:pfr} can be non-linear, it can
contain terms proportional to the field or to the field derivatives and it can
contain a non-vanishing constant. The latter might raise some concerns, as the
proof of the LSZ formula assumes a vanishing vacuum expectation value (vev) of
the operator $\Ocal$. Let us examine this issue. Suppose
$\delta Z[J]/\delta J^i(x)|_0= v^i$. If $v^i\neq 0$, it is customary to write
$\phi^i(x) = v^i+h^i(x)$ and work with the shifted fields $h^i$, which have
vanishing vev in the original theory $S$. Let
$\delta Z^\prime[J]/\delta J^i(x)|_0= \tilde{v}^i$. The corresponding shift is
$\phi^i(x) = \tilde{v}^i + \tilde{h}^i(x)$, such that $\tilde{h}^i$ has
vanishing vev in the theory $S^\prime$. The transformation $F$ induces another
transformation $\bar{F}$ on the shifted fields:
$h^i=\bar{F}^i(\tilde{h})=F^i(\tilde{v}+\tilde{h})-v^i = \tilde{h}^i + \lambda
\bar{G}^i(\tilde{h})$.  At the classical level, it can be easily checked that
$F^i(\tilde{v})=v^i$. This also holds at the quantum level when $F$ is
linear. In this case, $\bar{F}$ and $\bar{G}$ have no constant term.\footnote{This property is implicit in the discussion of spontaneously broken theories in~\cite{Arzt:1993gz}.} Conversely,
in this case the transformation
$\bar{F}^i(\tilde{h})=F^i(\tilde{v}+\tilde{h})-F^i(\tilde{v})$ leads to fields
$h^i$ with no vev. For generic non-linear transformations, on the other hand,
$F^i(\tilde{v}) \neq v^i$ at the quantum level. This can be seen as a particular
consequence of the fact that the quantum action (unlike the classical one) is
not a scalar under non-linear field redefinitions. This is due to the lack of covariance of the source terms
  $J_\alpha \phi^\alpha$: a non-linear field redefinition in this term cannot be
  absorbed into a redefinition of the sources. Covariant extensions of the effective action have been proposed in~\cite{Vilkovisky:1984st,Anselmi:2012jt}. 
  At any rate, in general $\bar{F}$ and $\bar{G}$ will have a constant at
$O(\hbar)$, and it is this constant that guarantees vanishing vevs. In practice,
all this amounts to performing a redefinition of fields with or without vevs, calculating the vevs of the new fields with the
new action and then performing the corresponding shifts (in perturbation theory
this can be achieved by imposing tadpole cancellation as a renormalization
condition, see the corresponding comments in~\cite{Passarino:2016saj}).%
\footnote{%
  Alternatively, it is possible to work with $h^\prime = F^{-1}(v+h)-F^{-1}(v)$,
  which in general will have a vev at $O(\hbar \lambda)$. The field $h^\prime$
  is perturbatively close to $\tilde{h}$ so the results will be the same in
  perturbation theory, although the presence of tadpoles is an unwanted
  complication. It can also be used in the LSZ formula, since the contribution
  of the (constant) difference with $\tilde{h}$ lacks the corresponding pole.%
} In particular, this means that in a theory with spontaneous symmetry breaking one can perform covariant field redefinitions in the symmetric phase, in such a way that the symmetry of the Lagrangian is kept manifest. In fact, this is a standard practice in the SMEFT. 

It should be remembered that the simplified result eq.~\refeq{eq:on-shell} is
not valid for off-shell quantities. We have already mentioned the fact that the
vevs of the fields are not covariant under field redefinitions. As pointed out
in~\cite{Passarino:2016saj}, care is also needed with unstable particles. Of
course, as long as they are rigorously treated as resonances in processes with
stable asymptotic states, the LSZ formula holds and eq.~\refeq{eq:on-shell} can
be used. The problem with eq.~\refeq{eq:on-shell} arises when one insists in
treating the unstable particles as external states, which is extremely useful
since most of the particles in the Standard Model (SM) are unstable. For this,
different treatments have been proposed (see ref.~\cite{Denner:2014zga} and
references therein).  It would be interesting to assess to what extent
eq.~\refeq{eq:on-shell} is a good approximation in each of these treatments.

To finish this section, let us discuss in what sense these results survive
renormalization. We can schematically write the generating function of
renormalized Green functions as
\begin{equation}
 \label{eq:renormalized-Z}
  Z^R[J]
  = \lim_{\eps\to 0}
  \int_{r_\eps} \Dcal\phi 
  \exp{
    \left(
      i S_\eps^R[\phi]
      + J_\alpha \phi^\alpha
    \right)},
\end{equation}
where $r_\eps$ represents some regularization with regulator $\eps$, which is
removed when taken to 0, while $S_\eps^R = S + S_{\eps}^{\mathrm{ct}}$, with
$S_{\eps}^{\mathrm{ct}}$ containing the necessary counterterms. Changing
variables again,
\begin{equation}
 \label{eq:transren-Z}
  Z^R[J]
  = \lim_{\eps\to 0}
  \int_{r^\prime_\eps} \Dcal\phi \det{
    \left( \frac{\delta F}{\delta \phi} \right) }
  \exp{
    \left(
      i S_{\eps}^R[F(\phi)]
      + J_\alpha F^\alpha(\phi)
    \right)
  },
\end{equation}
where $r^\prime$ is the regularization after the change of variables. This shows that
$(S^R)^{\prime}[\phi]=S^R[F(\phi)]$ with the regularization $r^\prime$ gives finite
Green functions of the operators $F(\phi)$ when $\epsilon\to 0$. Having in mind the simplified result eq.~\refeq{eq:on-shell} for on-shell
quantities, we are actually interested in the Green functions of fields $\phi$
in the theory $(S^R)^\prime$, but the substitution
$J_\alpha F^\alpha \to J_\alpha \phi^\alpha$ in \refeq{eq:transren-Z} produces a generating function that does not generate finite Green functions in the
limit $\eps \to 0$. Nevertheless, because of the on-shell equivalence with $S^R$, these Green functions will give finite scattering amplitudes (with the limit $\eps \to 0$ taken after the on-shell reduction). 

$S^\prime$ describes a (quasi) local field theory, so
it can also be renormalized (in a broad sense, possibly with infinite
counterterms), such that finite off-shell Green functions are found. The necessary counterterms and the corresponding
renormalized action $(S^{\prime})^R$ cannot be recovered by just making the same field
redefinition in the original renormalized action. That is,
$(S^{\prime})^ R \neq (S^{R})^{\prime}$. One nice way of relating the
renormalization in both theories
has been proposed in~\cite{Anselmi:2012aq}. The essential idea is to add sources
$L_a$ for all the possible operators. Then, it is shown that to connect both
renormalized theories not only the fields but also the sources must be
transformed: $\phi \to F(\phi)$, $L \to L^\prime(L)$. This is quite natural in
the framework of the renormalization of composite
operators~\cite{Shore:1990wp}, which is required here because
$\phi$ in the theory $S^\prime$ is composite from the point of view of the
original theory $S$. Interestingly, in this picture renormalization itself can
be seen as a regulator-dependent change of
variables~\cite{Anselmi:2012aq,Lizana:2017sjz}. The most important implication of these
relations between renormalized theories is that predictivity is preserved: if the
observables depend on a finite number of physical parameters, to a given order,
in the theory defined in the original variables, the same holds in the theory
defined with the new variables (see ref.~\cite{Bonneau:1985ea} for an explicit example in a renormalizable theory).

\section{Field redefinitions vs equations of motion}
\label{sec:eom}
The Schwinger-Dyson equations follow from the invariance of the path integral under infinitesimal field redefinitions\footnote{ Conversely, the path integral \refeq{eq:original-Z} can be understood as a formal
  solution to the Schwinger-Dyson equations. }
 and can be written succinctly as
\beq
\int \Dcal \phi \left[ i \frac{\delta S}{\delta \phi^\beta} + J^\beta \right] \exp (i S + J_\alpha \phi^\alpha ) = 0.
\label{eq:SD}
\eeq
Differentiation with respect to $J$ gives an infinite set of relations among the Green functions, which can be considered the quantum equations of motion of the theory. In this section, we discuss instead relations between field redefinitions and the {\em classical\/} equations of motion, $\delta S/\delta \phi^\alpha = 0$.

For the perturbative redefinition in eq.~\refeq{eq:pfr}, we can Taylor expand the resulting action,
\begin{align}
 S^\prime[\phi]  & =  S[F(\phi)] \nn
& =
 \sum_{m= 0}^\infty
 \frac{1}{m!} \lambda^{m}
 \,
 G^{\alpha_1}(\phi) \cdots G^{\alpha_m}(\phi)
 \frac{\delta^m S[\phi]}{\delta \phi^{\alpha_1} \ldots \delta \phi^{\alpha_m}} \nn
 & = S[\phi] + \lambda G^{\alpha_1}(\phi) \frac{\delta S[\phi]}{\delta \phi^{\alpha_1}} + O(\lambda^2) \nn
 &  =: S^\prime_{\mathrm{linear}}[\phi] + O(\lambda^2) .
 \label{eq:simpleexpansion}
 \end{align}
 The term linear in $G$, of order $\lambda$ is proportional to $\delta S/\delta \phi$, and thus vanishes if the classical equations of motion of $S$ are used. However, due to the higher-order terms, we see that $S'$ is not equal to $S^\prime_{\mathrm{linear}}$, that is, $S$ and $S^\prime_{\mathrm{linear}}$ are not related by this field redefinition for any $G$ and $\lambda$. As we show below, for a generic $G$ they are actually not related by any local field redefinition. Thus, adding to $S$ a perturbation proportional to its equations of motion does not result in general in an action equivalent to $S$. Equally, {\em eliminating terms in the action by imposing the classical equations of motion of the rest of the action does not produce an equivalent theory}. The equivalence only holds at linear order in the perturbation. Note that the perturbation $\lambda G \delta S/\delta \phi$ is neither redundant in the classical limit. Indeed, the relevant equations of motion for a tree-level calculation of Green functions include the variation of the perturbation itself and the variation of the source terms. 
  
 All this looks pretty straightforward, but apparently there is still some confusion about the limitations of the classical equations of motion, even among experts in effective theories. For example, statements such as ``the operators that vanish by the equations of motion are redundant'' or ``the operators that vanish by the equations of motion give no contribution to on-shell matrix elements'', without further qualification, are found every now and then in the specialized literature. 
To make this point completely clear, we stress that the proofs in~\cite{Politzer:1980me,KlubergStern:1975hc,GrosseKnetter:1993td,Wudka:1994ny} of the redundancy of equation-of-motion operators are only valid at the linear level, as indicated in these references. Let us briefly review the argument in~\cite{Politzer:1980me}, which is reproduced in the discussion about field redefinitions and equations of motion in the lecture notes~\cite{Manohar:2018aog}. Given an action $S$ and an operator of the form $\Ocal(z) = (f^i \delta S/ \delta \phi^i)(z)$, field redefinitions in the path integral are used to show that the correlators $\langle 0| T \phi^{i_1 x_1} \ldots \phi^{i_n x_n} \Ocal(z) |0 \rangle$ in the theory described by $S$ can be written as a sum of terms proportional to delta functions involving the points $i_1 \dots i_n$.\footnote{This is a simple generalization of the Schwinger-Dyson equations. Dimensional regularization is assumed in order to neglect the Jacobian of the transformation.}
Then, it follows from the LSZ formula that $\langle p_1 \ldots p_r| \Ocal(z)| p_{r+1} \ldots p_{n} \rangle$ vanishes, since the number of poles is smaller than $n$. From this, it is concluded in~\cite{Manohar:2018aog} that the operator $\Ocal$ ``can be dropped because it does not contribute to the S matrix''.  But this conclusion is an unjustified extrapolation of the particular result for S-matrix elements with only one insertion of $\Ocal$.\footnote{The author of ref.~\cite{Manohar:2018aog} warns latter that ``working to second order in the equations of motion is tricky'' (see ref.~\cite{Jenkins:2017dyc} for more details). However,  as shown in section~\ref{sec:perturbative} and in the example of section~\ref{sec:example}, using the equations of motion at second order is in general wrong, rather than tricky, while at first order it involves no complications.}   Indeed, the perturbative calculations with the complete action $S+\lambda \Ocal$ involve in general arbitrary powers of the interaction $\lambda \Ocal$, so one needs to also take into account the correlators $\langle 0| T \phi^{i_1 x_1} \ldots \phi^{i_n x_n} \Ocal(z_1) \ldots \Ocal(z_m) |0 \rangle$ with $m>1$. It can be checked that these correlators contain terms that are not proportional to any delta function involving the points $x_1,\dots, x_n$. These terms do not need to vanish when the elementary fields are reduced into on-shell particles. Therefore, diagrams with a single insertion of $\Ocal$ do not contribute when the external legs are on shell, but diagrams with two or more insertions do, in general.  In appendix~\ref{app:eom-counterexample}, we check explicitly in a simple example that, already at the tree level, $\langle p_1 p_2| T\Ocal(z_1) \Ocal(z_2)| p_3 p_4 \rangle \neq 0$. All this agrees with eq.~\refeq{eq:simpleexpansion}: $\lambda \Ocal$ can be eliminated at the linear order in $\lambda$ by a perturbative field redefinition, but in doing so other operators proportional to the second and higher powers of $\lambda$ are generated. The single (and multiple) insertions of these new operators reproduce the effect of the multiple insertions of $\Ocal$.

Another approach to the analysis of
redundancies in the action is focusing on {\em redundant parameters\/} instead of
redundant operators. In this case, there is an exact relation with the classical equations of motion. A parameter $\xi$ in an action $S_\xi$ will be redundant if it can be eliminated by a local field redefinition, i.e.\ if an invertible $F_\xi$ exists such that $S^\prime = S_\xi \circ F_\xi$ does not depend on $\xi$. Then, using that $\d S^\prime / \d \xi =0$, 
\begin{equation}
  \frac{\d S_\xi}{\d \xi}
  =
  \frac{\d (F_\xi^{-1})^\alpha}{\d \xi}
  \frac{\delta F_\xi^\beta}{\delta \phi^\alpha}
  \frac{\delta S_\xi}{\delta \phi^\beta}.  \label{eq:parameter-variation}
\end{equation}
We conclude that if $\xi$ is redundant, then $\d S_\xi/\d \xi$ vanishes when the classical equations of motion are enforced. The converse implication is also true: if  $\d S_\xi/\d \xi$ vanishes by the classical equations of motion, then $\xi$ is redundant~\cite{Weinberg:1995mt}. Indeed, the variation of $S_\xi$ under an infinitesimal change $\delta \xi$ of the parameter $\xi$ is $\delta S_\xi = (\d S_\xi/\d \xi) \delta \xi$. If $(\d S_\xi/\d \xi)=f^\alpha \delta S/\delta \phi^\alpha$, then the change $\delta S_\xi$ can be compensated by the infinitesimal transformation given by eq.~\refeq{eq:pfr} with $\lambda=\delta \xi$ and $G=-f$, as can be seen in eq.~\refeq{eq:simpleexpansion}. That is, $\d (S_\xi\circ F_\xi)/ \d \xi = 0$. Since this holds for any value of $\xi$, it follows that $S^\prime = S_\xi\circ F_\xi$ is constant in $\xi$.  

Let us use this last approach to study under which circumstances may $S+ \lambda f$ be equivalent to $S$, where $\lambda$ is a small parameter. Here, $f$ is a local functional of $\phi$ and both $S$ and $f$ may depend analytically on $\lambda$ at $\lambda=0$. Consider now the action $S + \xi f$, with $\xi$ a new parameter that is not present in $S$ nor in $f$. If $\xi$ were a redundant parameter, then the values $\xi=0$ and $\xi=\lambda$ would lead to equivalent actions, and we would conclude that $S+\lambda f$ is equivalent to $S$. 
As we have just seen, $\xi$ is a redundant parameter if and only if
\beq
\label{eq:redundant_eq}
f = g^\alpha \frac{\delta}{\delta \phi^\alpha} (S+\xi f).
\eeq
for some local $\xi$-dependent functionals $g^\alpha$ of $\phi$. We want to solve this equation for $g^\alpha$. To reduce the size of the equations, we use in the remaining of this section the notation $H_{,\alpha_1\dots \alpha_m} = \delta^m H / \delta \phi^{\alpha_1} \dots \delta \phi^{\alpha_m}$ for any functional $H$. As we are interested in perturbative redefinitions, we require that $g^\alpha$ has a power expansion $g^\alpha = g^{(0)\alpha} +  \xi g^{(1)\alpha} + \dots$.  Comparing the terms of order 0 in $\xi$, we see that, for a solution to exist, it must be possible to write $f$ in the form $f=f^\alpha S_{,\alpha}$, and then $g^{(0)\alpha}=f^\alpha$. Incidentally, this shows once more that the equations of motion can be employed to eliminate terms at first order; the necessary perturbative redefinition with $G^\alpha = -f^\alpha$ follows from eq.~\refeq{eq:parameter-variation}. Writing $g^\alpha =  f^\alpha + \xi \bar{g}^\alpha$, \refeq{eq:redundant_eq} reduces to
\beq
\label{eq:firstorder}
0 = \bar{g}^\alpha S_{,\alpha} + (f^\alpha+\xi \bar{g}^\alpha) (f^\beta_{,\alpha} S_{,\beta} + f^\beta S_{,\beta\alpha} )
\eeq
Looking at the leading order of this equation, we see that, for a solution to exist, we need
\beq
f^\alpha f^\beta S_{,\beta\alpha}  = h^\alpha S_{,\alpha},
\eeq
for some $h^\alpha$. For a non-trivial action $S$ and a generic $f^\alpha$, there is no solution to this equation, since the first and second derivatives give a non-homogeneous result when acting on terms in $S$ with different number of fields. A solution exists, however, if $f^\alpha = f^{\alpha \beta} S_{,\beta}$. Actually, in this case there is a solution of eq.~\refeq{eq:redundant_eq} to all orders in $\xi$,
since eq.~\refeq{eq:firstorder} is then of the form
\beq
0 = \left[ \bar{g}^\gamma + \left(f^{\alpha \beta} S_{,\beta} + \xi \bar{g}^\alpha \right) \left(f_{,\alpha}^{\gamma \delta} S_{,\delta}
+ 2 f^{\gamma \delta} S_{,\delta \alpha}\right)
 \right] S_{,\gamma} ,
\eeq
Thanks to its factorized form, this equation can always be solved recursively, looking for a local solution $\bar{g}^\alpha$ that makes the first factor (in brackets) vanish. This defines a local solution $g^\alpha$ as a power series in $\xi$. The local perturbative redefinition that eliminates $\xi f$ to all orders can then be obtained recursively, using eq.~\refeq{eq:parameter-variation}. Therefore, we conclude that the parameter $\xi$ is redundant in perturbation theory if $f$ is at least {\em quadratically} proportional to the equation-of-motion operator $\delta S/\delta \phi$. As anticipated, we can then conclude that the perturbation $\lambda f$ can be ignored on shell.
This result has actually been obtained before in~\cite{Anselmi:2006yh,Anselmi:2012jt}. Here, we have seen that for a general action this condition on $f$ is not only sufficient, but also necessary. A more direct way of checking that $\lambda S_{,\alpha} f^{\alpha\beta} S_{,\beta}$ is redundant is to perform a field redefinition to eliminate it at first order. Then, it is easy to check that the higher order terms have the same form. Therefore, successive field redefinitions will move the effects of the perturbation to higher and higher orders, while preserving the property that the generated terms are quadratic in the equation-of-motion operator. In this way, the effects of the perturbation can be completely eliminated up to an arbitrary power of $\lambda$.

\section{Field redefinitions and matching}
\label{sec:matching}

In this section we discuss the relation between field redefinitions performed
before and after the matching of an effective field theory to a UV completion, which may be another effective field theory. The
later will be called the fundamental or the UV theory. We assume that the action
$S_{\text{UV}}$ of the fundamental theory depends on a set of heavy fields
$\Phi$, with masses larger than a scale $\Lambda$ and a set of light fields
$\phi$, with masses lighter than $\Lambda$. An effective action $\bar{S}$ for
the light fields can defined by
\begin{align}
  Z[J]
  &=
  \int \Dcal \phi \Dcal \Phi
  \exp
  \left(
    i S_{\text{UV}}[\Phi,\phi]
    + J_\alpha \phi^\alpha
  \right)
  \nn
  &=
  \int \Dcal \phi
  \exp
  \left(
    i \bar{S}[\phi]
    + J_\alpha \phi^\alpha
  \right),
  \label{eq:matching}
\end{align}
with
\begin{equation}
  \exp
  \left(
    i \bar{S}[\phi]
  \right)
  =
  \int \Dcal \Phi
  \exp
  \left(
    i S_{\text{UV}}[\Phi,\phi]
  \right).
  \label{eq:effaction}
\end{equation}
If sources were added for the heavy fields, the effective action would also
depend on them. Without such sources, $Z[J]$ only generates the Green functions
of the light fields. The effective action $\bar{S}$ can be found by two
equivalent methods: A) requiring that the first and second lines
of eq.~\refeq{eq:matching} agree, which amounts to {\em matching} the off-shell 1PI
functions of the effective theory to the off-shell
one-light-particle-irreducible functions of the fundamental theory; B) {\em
  integrating out\/} the heavy degrees of freedom explicitly, i.e.\ computing
directly eq.~\refeq{eq:effaction}, for instance using functional methods.

The action $\bar{S}$, obtained by any of these methods, is non-local. However, a
local effective field action can be constructed to approximately reproduce the
function $Z$. The approximation is controlled by the dimensionful parameter
$\lambda=1/\Lambda$. Given a (non-local) effective action $\bar{S}$, we define
$\corrected{\bar{S}}{n}$ as the local action containing terms of order
$\lambda^n$ or less and such that
\begin{equation}
  \int \Dcal\phi
  \exp{
    \left(
      i \bar{S}[\phi] + J_\alpha \phi^\alpha
    \right)
  }
  =
  \int \Dcal\phi
  \exp{
    \left(
      i \corrected{\bar{S}}{n}[\phi] + J_\alpha \phi^\alpha
    \right)
  }
  + O(\lambda^{n+1}). 
  \label{eq:arreglo}
\end{equation}

The exact action $\bar{S}$ and generating function $Z$ can be viewed formally as
infinite series in $\lambda$:
\begin{equation}
  \bar{S}[\phi] = \sum_{k = 0}^\infty \lambda^k \bar{S}_k[\phi],
  \qquad
  Z[J]
  =
  \sum_{m = 0}^\infty
  \lambda^m Z_m[J].
\end{equation}
Note that each $\bar{S}_k$ is local, as adding derivatives to an operator
increases its order in $\lambda$.  But because $\bar{S}$ is non-local, it turns
out that knowing $\bar{S}_k$ for $k \leq n$ for any given $n$ is not enough, in
general, to compute $Z_n$ at the quantum level. Even if any sum of a finite number of terms with
$k > n$ gives a vanishing contribution to $Z_n$, the tail
$\sum_{k \geq N} \lambda^k \bar{S}_k$ of the series may contribute to it for
arbitrarily large $N$. Therefore, the naive truncation $\truncated{\bar{S}}{n}$
of $\bar{S}$ to order $n$ does not coincide, in general, with the local
effective action $S = \corrected{\bar{S}}{n}$, which gives the correct
approximation of $Z$ to order $n$.

In the saddle-point expansion, this can be understood in the following
way~\cite{delAguila:2016zcb}. Whenever the linear couplings of the heavy fields to the light ones are non-vanishing, 
the exact saddle-point configuration, which gives the
effective action at the tree level, needs to be approximated by a truncated
expansion in $\lambda$, say to order $N$. Then, besides the usual quadratic and
higher-order terms, the heavy-field expansion of the action about this non-exact
saddle point includes linear terms suppressed by $\lambda^{N+1}$ and higher
powers of $\lambda$. Despite this suppression, at the quantum level such terms must be taken into
account in the integral of the heavy fields $\Phi$. Indeed, the quantum
corrections may give contributions to orders $k < N$, independently of how large
$N$ is.  The essential reason is that loop integrals regularized with
dimensionless regulators probe all energy scales, including those higher than
$\Lambda$.  A way of finding these contributions within approach B, based on the method of regions~\cite{Beneke:1997zp}, has been
proposed in~\cite{Fuentes-Martin:2016uol}. In the matching approach A,
correcting $\truncated{\bar{S}}{n}$ to find $\corrected{\bar{S}}{n}$ is not
really an issue: in practice, the matching is performed directly between
$S_{\text{UV}}$ and $S=\corrected{\bar{S}}{n}$ at some given $n$. The necessary
contributions then appear automatically from diagrams in the fundamental theory
involving loops of both light and heavy
fields, which at one loop are only present when the heavy fields have linear couplings to the light ones~\cite{Bilenky:1993bt,Bilenky:1994kt,delAguila:2016zcb}.

Once again, a renormalization procedure is required to make sense of all these
equations. The matching can be performed at the regularized level (with the same
dimensionless regulator). This leads to a regularized effective action that can
be perturbatively renormalized. But it makes more physical sense to match the
renormalized theories, as at the end of the day the aim of matching is to
express the renormalized parameters of the local effective action $S$ as
functions of the renormalized parameters of $S_{\mathrm{UV}}$. In method B, this
can be achieved by adding counterterms to the UV action but refraining from
removing the regulator; then the necessary counterterms in the effective theory
will be generated (in the same regularization and renormalization scheme) during
the matching procedure. The UV behaviours of the fundamental and effective
theories are different, and so will be the counterterms. In the standard
approach to matching within method A, the renormalized Green functions of the
fundamental and effective theories are compared (with removed regulators).  This
allows great flexibility, as neither the regularization method nor the
renormalization scheme need to be the same in both theories. The relation
between renormalized parameters depends on these schemes. To preserve this
relation, the effective theory should be used in the same scheme used for
the matching. In this regard, observe that, because the effective theory is
local, all its renormalized couplings and masses can be modified by finite
counterterms. Hence, by adapting the scheme to each UV theory, all the UV
information in the renormalized parameters of the effective theory can be
erased. Scheme independence, however, ensures that the calculations done in such
a scheme (which will depend on the UV parameters) will reproduce to the required order the low-energy
predictions of the corresponding fundamental theory. In practice, it is
preferable to see this information explicitly in the renormalized parameters, so
a universal renormalization scheme, such as MS, should be used in the effective
theory.

Let us now perform a general local change of variables involving both the heavy
and the light fields,
$(\Phi,\phi) \to F(\Phi,\phi) = (F_h(\Phi,\phi), F_l(\Phi,\phi))$. We find
\begin{equation}
  Z[J] =
  \int \Dcal \phi \Dcal \Phi
  \det
  \left(
    \frac{\delta F}{\delta (\phi,\Phi)}
  \right)
  \exp
  \left\{
    i S_{\text{UV}}[F(\Phi,\phi)]
    + J_\alpha F_l^\alpha(\Phi,\phi)
  \right\}.
\end{equation}
Consider first the particular case with $F_l(\Phi,\phi) = F_l(\phi)$, that is to
say, the case in which the new light fields depend only on the original light
fields.%
\footnote{%
  This kind of redefinition is implicitly performed in the method proposed in
  ref.~\cite{Henning:2016lyp} to account for the heavy-light loop contributions.
}
Then,
\begin{align}
  Z[J]
  &=
  \int \Dcal \phi \Dcal \Phi
  \det
  \left(
    \frac{\delta F_l}{\delta \phi}
  \right)
  \det
  \left(
    \frac{\delta F_h}{\delta \Phi}
  \right)
  \exp
  \left\{
    i S_{\text{UV}}[F_h(\Phi,\phi),F_l(\phi)]
    + J_\alpha F_l^\alpha(\phi)
  \right\}
  \nn
  &=
  \int \Dcal \phi
  \det
  \left(
    \frac{\delta F_l}{\delta \phi}
  \right)
  \exp
  \left\{
    \overline{S^\prime}[\phi]
    + J_\alpha F_l^\alpha(\phi)
  \right\},
  \label{eq:int1}
\end{align}
where
\begin{align}
  \exp
  \left(
    i \overline{S^\prime}[\phi]
  \right)
  &=
  \int \Dcal \Phi
  \det
  \left(
    \frac{\delta F_h}{\delta \Phi}
  \right)
  \exp
  \left\{
    i S_{\text{UV}}[F_h(\Phi,\phi), F_l(\phi)]
  \right\}
  \nn
  &=
  \int \Dcal \Phi
  \exp
  \left\{
    i S_{\text{UV}}[\Phi, F_l(\phi)]
  \right\}.
  \label{eq:int2}
\end{align}
In the last line we have redefined back the heavy variables for fixed light
fields. This change of variables is given by
$\Phi = F^{-1}_h(\Phi', F_l(\phi))$, with $F^{-1}_h$ defined by
$F^{-1}(\Phi, \phi) = (F^{-1}_h(\Phi, \phi), F^{-1}_l(\Phi, \phi))$.

The last equation shows that
\begin{equation}
  \label{eq:simplematching}
  \overline{S^\prime}[\phi] = \bar{S}[F_l(\phi)], 
\end{equation}
which is also consistent with a change of variables in eq.~\refeq{eq:effaction}. So,
for the transformations we are considering now, the heavy field redefinition
does not modify $\bar{S}$, while the light field redefinition commutes with the
integration of the heavy field.

However, the local version of eq.~\refeq{eq:simplematching}, \beq
\corrected{\overline{S^\prime}}{n}[\phi] \overset{?}{=}
\corrected{\bar{S}}{n}[F_l(\phi)], \label{eq:nofunciona} \eeq does not hold, in
general. Here, both $\corrected{\bar{S}}{n}$ and
$\corrected{\overline{S^\prime}}{n}$ are defined by eq.~\refeq{eq:arreglo} (with
sources coupling linearly to $\phi$).  Eq.~\refeq{eq:nofunciona} is equivalent
to
\begin{align}
  \label{eq:tilde-redefined-sources}
  \int \Dcal\phi
  \exp{
    \left(
      i \bar{S}[\phi]
      + J_\alpha {(F_l^{-1})}^\alpha(\phi)
    \right)
  }
  \overset{?}{=}
  \int \Dcal\phi
  \exp{
    \left(
      i \corrected{\bar{S}}{n}[\phi]
      + J_\alpha {(F_l^{-1})}^\alpha(\phi)
    \right)
  }
  +
  O(\lambda^{n + 1}),
\end{align}
as can be seen by performing a redefinition $\phi \to F_l(\phi)$, using the
definition of $\corrected{\overline{S^\prime}}{n}$, its assumed equality with
$\corrected{\bar{S}}{n} \circ F$ and performing another redefinition
$\phi \to F_l^{-1}(\phi)$. But requiring agreement to a given order of the Green
functions of $\phi$ is not the same as requiring agreement to that order of the
Green functions of $F_l^{-1}(\phi)$. To prove this, we present in
appendix~\ref{app:simple-matching-counterexample} a counterexample, which shows
that eq.~\refeq{eq:tilde-redefined-sources} is not satisfied in general at the
quantum level. Of course, even if the redefinition $F$ does not convert
$\corrected{\bar{S}}{n}$ into $\corrected{\overline{S^\prime}}{n}$, these two
actions are equivalent on shell.

All this discussion applies irrespectively of whether method $A$ or $B$ is
employed for the matching. Let us add a few remarks on method A. In this method,
the matching is standardly performed for Green functions of the fields $\phi$
that appear in the action, be it the original or the transformed one. If the
comparison with the Green functions for action $S_{\mathrm{UV}}$ or
$S_{\mathrm{UV}}^\prime$ is performed with a general local effective action that
includes all the symmetric operators to a given order, then $S$ or $S^\prime$
will be automatically found, respectively. As we have shown, they will be
equivalent, but not directly related by the transformation $F$. A problem may
arise if a non-redundant basis is employed. Then it is not possible, in
general, to adjust the coefficients in such a way that the off-shell Green
functions reproduce those of the fundamental theory with an arbitrary
$S_{\mathrm{UV}}$. Indeed, proceeding in this way would be like trying to match
Green functions of different fields, $\phi$ and
$\phi^\prime=F(\phi)$. Therefore, any conversion into a reduced basis
should be performed after the (off-shell) matching, also in method A. The
alternative is to require only agreement for on-shell quantities, as proposed
in~\cite{Georgi:1991ch}.

In eq.~\refeq{eq:int1} and \refeq{eq:int2} we have used in several places
(determinant, action and source terms) the fact that $F_l$ is independent of
$\Phi$. Therefore, the simple relation eq.~\refeq{eq:simplematching} cannot be
extended to the general case in which $F_l$ depends on the heavy fields.%
\footnote{%
  The redefinition used in the method of ref.~\cite{Fuentes-Martin:2016uol} to
  account for heavy-light loops belongs to this more general case.%
}
Nevertheless, as long as the redefined light field is a valid interpolating
field for the light particles, we have
\begin{align}
  Z[J]
  &\sim
  Z^\prime[J]
  \nn
  &=
  \int \Dcal \phi \Dcal \Phi
  \det
  \left(
    \frac{\delta F}{\delta (\phi,\Phi)}
  \right)
  \exp
  \left\{
    i S_{\text{UV}}[F(\Phi,\phi)]
    + J_\alpha \phi^\alpha
  \right\}
  \nn
  &=
  \int \Dcal \phi
  \exp
  \left\{
    i \bar{S}^{\prime\prime}[\phi]
    + J_\alpha \phi^\alpha
  \right\},
\end{align}
with
\begin{equation}
  \exp
  \left(
    i \bar{S}^{\prime\prime} [\phi]
  \right)
  =
  \int \Dcal \Phi
  \det
  \left(
    \frac{\delta F}{\delta (\phi,\Phi)}
  \right)
  \exp
  \left\{
    i S_{\text{UV}}[F(\Phi,\phi)]
  \right\}.
\end{equation}
$\overline{S^{\prime\prime}}$ (and the corresponding
$\corrected{\overline{S^{\prime\prime}}}{n}$) can be used to compute on-shell
amplitudes of light particles, even if it has no general simple connection with
$\bar{S}$ ($\corrected{\bar{S}}{n}$).

In addition to these remarks, note that the discussion in the previous section
about renormalization before and after the field redefinition also applies to the
fundamental and effective renormalized theories that enter the matching.

\section{Perturbative expansions}
\label{sec:perturbative}

\subsection{Removing reparametrization redundance}
\label{subsec:nonredundant}

The theory space of possible actions with a given field content can be divided
into equivalence classes, with actions in the same class related by field
redefinitions (possibly with some restrictions, as discussed in
section~\ref{sec:reparametrization-invariance}).  All the actions in the same
class give rise to the same S matrix. An elegant way of working with these equivalent classes, which has been mostly employed in non-linear sigma models, is to use a geometric approach, in which the fields are coordinates of a differentiable manifold with a connection~\cite{Meetz:1969as,Honerkamp:1996va,Honerkamp:1971sh,Ecker:1972bm,AlvarezGaume:1981hn,AlvarezGaume:1981hm,Boulware:1981ns,Vilkovisky:1984st,Alonso:2016oah}. This allows to maintain explicit covariance under changes of coordinates (that is, field redefinitions). Here we will study the more mundane (but also useful) approach of 
choosing a representative for each equivalence class and
systematically reducing every action to the corresponding representative~\cite{Coleman:1969sm}. This is what we called ``fixing a gauge'' in the introduction. In this subsection, we first review how this gauge fixing can be performed order by order in perturbation theory and then examine the consequences of this procedure. 

The effective field theory is organized as a power series in $\lambda=1/\Lambda$:
\begin{equation}
  \label{eq:expansion}
  S[\phi] = \sum_{n = 0}^\infty \lambda^n S_n[\phi]. 
\end{equation}
Let us study the effect of local perturbative redefinitions of order $k$, of
the form $F(\phi) = \phi + \lambda^k G(\phi)$, with $k\geq 1$ and $G$ analytic
in $\lambda$.  Under this redefinition, the action changes into
\begin{align}
  \label{eq:expanded-action}
  S^\prime[\phi] & = S[F(\phi)] \nn
  &=
  \sum_{n, m = 0}^\infty
  \frac{1}{m!} \lambda^{n + k m}
  \,
  G^{\alpha_1}(\phi) \cdots G^{\alpha_m}(\phi)
  \frac{\delta^m S_n}{\delta \phi^{\alpha_1} \ldots \delta \phi^{\alpha_m}}
  \nn
  &=
  S[\phi]
  + \lambda^k G^\alpha(\phi) \frac{\delta S_0}{\delta \phi^\alpha}
  + O\left(\lambda^{k + 1}\right).
\end{align}
In particular, the last line of this equation shows that all the actions that
differ by order-$k$ terms proportional to the lowest-order equation of motion
belong to the same class to order $k$.  Suppose $S_k$ contains a term of
the form $f_k^\alpha(\phi) \delta \mathcal{K} / \delta \phi^\alpha$, with
$\mathcal{K}$ any term in $S_0$. Then, this term can be eliminated by the
following field redefinition of order $k$: \beq F_k^\alpha(\phi) = \phi^\alpha -
\lambda^k f_k^\alpha(\phi) .  \label{eq:redundant-redef} \eeq Obviously, this
redefinition has no effect to order $k-1$. At order $k$, its only effect is to add
$-f^\alpha_k  \delta S_0 / \delta \phi^\alpha$ to the action, which is the
same as using the lowest-order equation of motion to change
$ \delta \mathcal{K} / \delta \phi^\alpha$ by
$\delta (\mathcal{K}-S_0) / \delta \phi^\alpha$. The
redefinition eq.~\refeq{eq:redundant-redef} also changes the action at order $k+1$
and higher, as indicated in eq.~\refeq{eq:expanded-action}.

Therefore, once the lowest order action $S_0$ is fixed, a representative of each equivalence class can be chosen, order by order,
by picking at each order $k\geq 1$ a specific term $\mathcal{K}_k$ (which could be a linear combination of other terms) of $S_0$ and imposing (besides the hermiticity of the action and invariance under the relevant symmetries) that the coefficients of operators in
$S_k$ proportional to $\delta \mathcal{K}_k / \delta \phi^\alpha$ be equal to
zero. Identifying these operators may require algebraic manipulations and integration by parts. Note that, for a given $\mathcal{K}_k$, the maximal number of different factors $\delta \mathcal{K}_k / \delta \phi^{ix}$ is equal to the number of different fields $\phi^i$. Therefore, to eliminate all the ambiguities at each order $k$, $\mathcal{K}_k$ should be chosen such that $\delta \mathcal{K}_k / \delta \phi^{ix}\neq 0$ for all $i$.  A standard choice that works for any $k$ is to take $\mathcal{K}_k$ as the sum of all the kinetic terms.
Then, any subsequent redefinition of order $k$ would move the action into a different gauge, so the remaining linearly-independent operators that can appear in $S_k$ will form a non-redundant
basis of operators at that order.  To reach this basis from
an arbitrary effective action, one proceeds order by order. Let $S^{(k-1)}$ be
the transformed action after consecutive field redefinitions
$F_1,\dots, F_{k-1}$ that put it in the prescribed form to $O(\lambda^{k-1})$
and let $f_k(\phi)$ be the coefficient of
$\delta \mathcal{K}_k / \delta \phi^\alpha$ in $S^{(k-1)}$. Then, the field
redefinition eq.~\refeq{eq:redundant-redef} transforms $S^{(k-1)}$ into $S^{(k)}$,
which is in the prescribed form to $O(\lambda^k)$.  The actions $S^{(k)}$ and
$S$ are connected by the field redefinition
$F=F_k \circ F_{k-1} \circ \cdots \circ F_1$.

We see that, in order to define a non-redundant basis of operators, it is enough to use the lowest-order equations of motion in the operators to be eliminated~\cite{Georgi:1991ch}. Indeed, for this purpose, and as long as all the
algebraically-linearly-independent operators are included from the very beginning, the higher order
corrections at each step $k$ are absorbed into coefficients that were
arbitrary anyway, so there is no need to worry about
them. In fact, the same holds for the coefficients of the non-vanishing operators at order $k$. So, as described in the last paragraph, it actually suffices to identify a set of appropriate $\mathcal{K}_k$ and put to zero all the terms proportional to $\delta \mathcal{K}_k/\delta \phi^\alpha$. However, we have already stressed that it is often important to know the dependence of the coefficients in the transformed action on the original ones. Then, the redefinition must be performed explicitly. When working to next-to-leading order, $n=1$, the algorithm has only one step ($k=1$) and it is sufficient to apply the equations of motion of $S_0$ to the operators to be eliminated. But when working at orders
$n\geq 2$, it is mandatory to include the higher-order corrections in the
redefinition. This is the case when one wants to rewrite a known action $S$ in
a particular basis. To second order, for instance, this can always be achieved
as explained above by a field redefinition $F= F_2 \circ F_1$, where
$F_k^\alpha(\phi) = \phi^\alpha + \lambda^k G^\alpha_k(\phi)$, with $G_k$ a 
$\lambda$-independent function of the parameters of $S_m$ , $m\leq
k$. The redefined action is
\begin{align}
  S^\prime[\phi]
  &=
  S[F(\phi)]
  \nn
  &=
  S_0[\phi]
  + \lambda
  \left[
    S_1
    + G_1^\alpha(\phi) \frac{\delta S_0}{\delta \phi^\alpha}
  \right]
  \nn
  &
  ~~~ \mbox{}
  + \lambda^2
  \left[
    S_2
    + G_1^\alpha(\phi) \frac{\delta S_1}{\delta \phi^\alpha}
    + \frac{1}{2} G_1^\alpha(\phi) G_1^\beta(\phi)
    \frac{\delta^2 S_0}{\delta \phi^\alpha \delta \phi^\beta}
    + G_2(\phi) \frac{\delta S_0}{\delta \phi^\alpha} \right]
  + O(\lambda^3)
  \nn
  &=
  S^\prime_0[\phi] + \lambda S^\prime_1[\phi] + \lambda^2 S^\prime_2[\phi] + O(\lambda^3).
  \label{eq:changebasis}
\end{align}
We see explicitly that $S^\prime_{k}$ depends in general on the parameters of
all $S_n$ with $n\leq k$, and also that the higher-order effect of $F_1$ must be
taken into account in order to get the correct dependence of the parameters of
$S^\prime_2$ on the parameters of $S_0$, $S_1$ and $S_2$.  In particular,
\refeq{eq:changebasis} is relevant when comparing, to second order in $\lambda$,
the constraints on the operator coefficients in one basis with the ones in
another basis.  The same considerations apply to perturbative matching: {\em
  field redefinitions performed to eliminate terms of order $k$ in the effective
  action have an impact on the matching not only at order $k$ but also at higher
  orders.} This is readily seen in eq.~\refeq{eq:redundant-redef}
and eq.~\refeq{eq:changebasis}, taking $S$ to be the local effective action obtained from
matching to a more fundamental theory. For instance, even if we put $G_2=0$
in eq.~\refeq{eq:changebasis}, we cannot say that to order 2 this $S$ is equivalent
to $S_0+S_1^\prime+S_2$. Changing $S_1$ by $S_1^\prime$ requires in general a
change $S_2 \to S_2^\prime$. Observe also that 
knowledge of the field redefinition $F$ (in particular of $F_1$) is needed to find the correct $S_2^\prime$.

We stressed in section~\ref{sec:eom} that using the exact classical equations of motion is not equivalent to a non-infinitesimal field redefinition, and that it does not lead to an equivalent action. Perturbatively, the effect of an order-$k$ field
redefinition can be written as
\begin{equation}
  S^\prime[\phi] = 
  S[\phi]
  + \lambda^k \, G(\phi)^\alpha \frac{\delta S}{\delta \phi^\alpha}
  + O\left(\lambda^{2 k}\right).
\end{equation}
The exact equations of motion only give the linear contribution, starting at $\lambda^k$, but miss the remaining $O(\lambda^{2k})$ terms, which are necessary for $S^\prime$ to be equivalent to
$S$.  Hence, the equations of motion at higher orders, as used for instance in~\cite{Jenkins:2017dyc,Barzinji:2018xvu},
are not sufficient to find the higher-order corrections induced by a field
redefinition. In
particular, using  in $S_1$
(and $S_2$) the equation of motion to second order in $\lambda$ does not give, in general, an action that is equivalent to $S$ to
second order. The same conclusions apply to the case in which the equations of motion of $S^\prime$ are used in $S_1$ (and $S_2$), as can be seen by exchanging the roles of $S$ and $S^\prime$ and considering the inverse transformation.
To obtain the correct $S^\prime$, it is necessary to perform the
actual field redefinition in every term of the original action. This can be done
either directly or using the functional-derivative expansion in the second line of eq.~\refeq{eq:expanded-action}.

\subsection{Power counting}
\label{subsec:powercounting}

The effective field theories of interest often depend on several parameters,
which can be taken to be the cutoff scale $\Lambda$ and additional dimensionless
quantities, such as coupling constants, ratios of masses and $4\pi$ factors
associated to loops. The effective theory is organized as a multiple power
series in $1/\Lambda$ and certain combinations of the parameters, which are
assumed to be small (compared to the probed energies, if dimensionful). In the
following we use $\eta$ to refer to $1/\Lambda$ and any of these
combinations. For example, chiral perturbation theory is arranged as a power
series in $1/\Lambda$ with $\Lambda=4\pi f$ and $f$ the pion decay constant. One
could consider a simultaneous expansion in $1/f$ at each order in $1/\Lambda$,
but this expansion is conveniently resummed using the underlying structure of an
spontaneously broken theory. To organize systematically these expansions, it is
important to have a power-counting rule that assigns a number $N_\eta(\Ocal)$ to
each operator $\Ocal$ and each parameter $\eta$. Then, the ``natural''
coefficient of an operator $\Ocal$ is given by
\begin{equation}
  C_\Ocal \simeq \prod_\eta \eta^{N_\eta(\Ocal)}.
\end{equation}
For instance, in chiral perturbation theory, chiral counting dictates that
$N_{1/\Lambda}(\Ocal)$ is equal to the number of derivatives in $\Ocal$. In some
cases it is convenient to include in the specification of the operator not only
fields and derivatives but also powers of particular coupling constants or masses, which are treated as spurions and taken into account in the counting. To
guarantee the stability of the loop expansion, the power-counting rule should be
such that all the diagrams that can generate an operator give a contribution
that is similar to or smaller than its natural coefficient. In particular, this
requires
\begin{equation}
  \label{eq:factorize}
  \Delta_\eta(\Ocal_1 \Ocal_2) = \Delta_\eta(\Ocal_1)+\Delta_\eta(\Ocal_2) , 
\end{equation}
where $\Delta_\eta(\Ocal) = N_\eta(\Ocal) + c_\eta$ for some $c_\eta$ independent of the operator. 
A power-counting rule that is appropriate in many circumstances and enjoys nice
properties is naive dimensional analysis
(NDA)~\cite{Manohar:1983md,Jenkins:2013sda,Gavela:2016bzc}. In this case
the actual numerical coefficients are expected to be approximately equal to
their natural values when the UV completion is strongly coupled, and smaller
than them when it is weakly coupled. Approximate symmetries or tunings in the
fundamental theory can also give rise to smaller coefficients. Certain
assumptions on the UV theory allow to incorporate these suppressions
systematically in the power-counting rules~\cite{Giudice:2007fh}.

Now, we would like to examine the order of the new terms generated by field redefinitions, according to the assumed power counting, and to determine the conditions for their coefficients to be natural. 
Consider a field redefinition given by
$F^\alpha(\phi) = \phi^\alpha + f^\alpha(\phi)$, with $f$ local. We can write
each $f^j$ as a linear combination of local operators
$f^j_1, \ldots, f^j_{n_j}$. The redefinition will be perturbative when
$\mathrm{min} \{N_\eta(f^j_1),\ldots,N_\eta(f^j_{n_j}) \} >
N_\eta(\phi^j)$ for some $\eta$. Thanks to the factorization property
eq.~\refeq{eq:factorize}, the redefinition preserves the counting rule: an
action with natural operator coefficients is transformed into an action with
natural operator coefficients whenever the coefficients of the operators in $f$
are natural. The latter means that the coefficient $\alpha^j_i$ of each operator
$f^j_i$ is
\begin{equation}
  \label{eq:natural}
  \alpha^j_i \simeq \prod_\eta
  \eta^{\Delta_\eta(f_i^j)-\Delta_\eta(\phi^j)} = \prod_\eta
  \eta^{N_\eta(f_i^j)-N_\eta(\phi^j)} .  
\end{equation}
This condition will always be satisfied if the redefinition is performed to
eliminate any term in an action with natural coefficients.%
\footnote{%
  If the definition has any other purpose, coefficients $\alpha^j_i$ smaller or
  larger than \refeq{eq:natural} (that is, ``under-natural'' and
  ``super-natural'', respectively) are possible that still preserve the perturbativity
  of the transformation. Super-natural coefficients will give rise to
  perturbative corrections that destabilize the hierarchical structure of the
  original effective action. This will not be reflected in on-shell quantities
  if all the new terms are included, since the new action is equivalent to the
  original one. But the perturbative orders will be mixed, which must be taken
  into account in truncations of the new action.%
}
Explicitly, if the redefinition removes a term
$\mathcal{Q} = f^\alpha (\delta \mathcal{K}/\delta \phi^\alpha)$, with
$\mathcal{K}$ any term in the original action $S$, an operator $\Ocal$ in the
original action will give rise to a sum of terms of the form
\begin{equation}
\label{eq:Om}
  \Ocal_{[m]}
  =
  f^{\alpha_1}\dots f^{\alpha_m} \frac{\delta^m \Ocal}{\delta \phi^{\alpha_1} \dots \delta \phi^{\alpha_m}},
\end{equation}
with power counting given by
\begin{equation}
  \label{eq:power-counting}
  N_\eta(\Ocal_{[m]})
  =
  N_\eta(\Ocal)
  + m \cdot (N_\eta(\mathcal{Q}) - N_\eta(\mathcal{K})).
\end{equation}
We have used the factorization property
$ \Delta_\eta(\mathcal{Q}) = \Delta_\eta(f) + \Delta_\eta(\mathcal{K}) - \Delta_\eta(\phi) $. If the
coefficient in $\mathcal{Q}$ happens to be suppressed by a factor $\xi$,
relative to its natural value, while $\Ocal$ has a coefficient suppressed by a
factor $\kappa$ and $\mathcal{K}$ is natural, then the contribution $\Ocal_{[m]}$ in
$S^\prime$ will be be suppressed by $\xi^m \kappa$.

In the rest of the section we point out a few implications of this counting when
working with the SMEFT~\cite{Brivio:2017vri}. This effective theory is usually described as having a power
counting determined by the canonical dimension $\Delta$ of the operators:
$N_{1/\Lambda}(\Ocal)= \Delta(\Ocal)-4$. In this case, $\Delta_{1/\Lambda}=\Delta$ and $c_{1/\Lambda}=4$. We ignore in the following the few operators of dimension 5 and 7. In order to reach some standard basis
at dimension 6, one may need to redefine the Higgs doublet $\phi$ in such a way
that the dimension-6 terms proportional to $\Box \phi$ are removed. The
necessary cancellation arises from the kinetic term, while the remaining terms
of dimension 4 generate other terms (of the form $\Ocal_{[1]}$ in eq.~\refeq{eq:Om}) at
dimension-6. This is the same as using the dimension-4 Higgs equation of motion
in the terms to be eliminated. As discussed in the previous section, there will
be corrections at dimension 8, from substituting one $\phi$ at dimension 6 or
two $\phi$ at dimension 4. Note, however, that an important detail is missing in
this discussion: the SMEFT does not start at dimension 4. The gauge-invariant
operator $\Ocal_\mu=\phi^\dagger \phi$ has canonical dimension 2. Under the same
field redefinition, this super-renormalizable operator gives contributions of
the form $(\Ocal_\mu)_{[1]}$ of dimension 4. Even if one can absorb the corrections into a
renormalization of the SM couplings, this renormalization modifies
the coefficients at dimension 6~\cite{deBlas:2017xtg}. These linear contributions can
also be found using the equations of motion. But on top of this, $\Ocal_\mu$
contributes at dimension 6 with terms of the form $(\Ocal_\mu)_{[2]}$. Indeed, using eq.~\refeq{eq:power-counting} in this particular
case, we find $N_{1/\Lambda}((\Ocal_\mu)_{[2]}) = -2 + 2 \cdot 2 =2$. Such a generated term can be seen explicitly in the example of section~\ref{sec:example}. Because $(\Ocal_\mu)_{[2]}$
is proportional to $\delta^2 \Ocal_\mu/\delta \phi^2$, these dimension-6
contributions will be missed if one only uses the equations of motion. Note that
this does not contradict the standard procedure to reach a basis by using the
equations of motion, reviewed in section~\ref{sec:perturbative}, because the action at
leading order is not given by the dimension-4 terms but by the integral of
$\tilde{\Ocal}_\mu = -\mu^2 \Ocal_\mu$. Thus, the field redefinition we are
considering has nothing to do with the equations of motion of the action at
leading order.%
\footnote{%
  The equation of motion at leading order is just $\phi=0$. This could be used
  to eliminate recursively all the terms containing the Higgs doublet at
  dimension 4 and above. This looks strange, but it is consistent with the
  natural value of $\mu^2$ being of order $\Lambda^2$, according to the
  dimensional counting. Actually, a field with a mass of the order of the cutoff
  will decouple from the other fields. More precisely, it should be integrated
  out.%
}

Of course, the coefficient $\mu^2$ of $\Ocal_\mu$ is not natural with the counting
based on dimensions. Experimentally, we know that there is a hierarchy
$\mu \ll \Lambda$.  Hence, the new terms $(\Ocal_\mu)_{[1]}$ and $(\Ocal_\mu)_{[2]}$ arising
from $\Ocal_\mu$ will carry an extra suppression $(\mu/\Lambda)^2$ and will
typically be less important, numerically, than the corresponding dimension-4 and dimension-6 terms. This can
be rephrased in a more systematic way by incorporating $\mu$ in the power
counting: $\Delta_{1/\Lambda}(\mu^2) = 2$. This modified counting is nothing but
dimensional analysis. It follows that $N_{1/\Lambda} (\tilde{\Ocal}_\mu) =
0$. So, with the new counting $\tilde{\Ocal}_\mu$ is of the same order as the
dimension-4 terms, and the SM is the leading order approximation of the
SMEFT.

Consider next (differential) cross sections calculated in the SMEFT to order $n$
in $1/\Lambda^2$. They are schematically of the form
\begin{equation}
  \label{eq:cross-section}
  \sigma
  \propto
  {
    \left|
      A^{(0)}
      + \frac{1}{\Lambda^2} A^{(1)}
      + \frac{1}{\Lambda^4} A^{(2)}
      + \cdots
    \right|
  }^2,
\end{equation}
where $A^{(n)}$ is the coefficient of $\Lambda^{-2n}$ in the $1/\Lambda^2$
expansion of the on-shell amplitude. We denote by $A^{(n)}_{i_1 i_2 \ldots i_k}$
the part of $A^{(n)}$ given by diagrams with $k$ insertions of operators, one
from $S_{i_1}$, another one from $S_{i_2}$, etc. Then, we have:
\begin{equation}
  A^{(n)} = \sum_{i_1 + i_2 + \cdots + i_k = n} A^{(n)}_{i_1 i_2 \ldots i_k}.
\end{equation}

We ignore here phase-space factors, as we are only going to discuss the relative
importance of the quadratic and interference terms in the evaluation of the
right-hand side of eq.~\refeq{eq:cross-section}, which are schematically of the form
$A^{(i)} A^{(j)}$, with $i,j \leq n$. Let us nevertheless refer
to~\cite{Gavela:2016bzc} for an interesting result for the scaling of total
cross sections in NDA. Expanding eq.~\refeq{eq:cross-section},
\begin{equation*}
  \sigma
  \propto
  {\left| A^{(0)} \right|}^2
  + \frac{2}{\Lambda^2} \re \left(A^{(0)*} A^{(1)}_1\right)
  + \frac{1}{\Lambda^4}
  \left[
    {\left| A^{(1)} \right|}^2
    + 2 \re
    \left(
      A^{(0)*} A^{(2)}_{11}
      + A^{(0)*} A^{(2)}_2
    \right)
  \right]
  + O\left(\frac{1}{\Lambda^6}\right),
\end{equation*}
where we have grouped contributions of the same order. In many applications,
only the first two terms need to be taken into account. However, there are
processes in which the interference terms $\re (A^{(0)*} A^{(1)})$ vanish (or
are very suppressed)~\cite{Helset:2017mlf}. Then the terms in brackets give the leading correction and
must be included in the analysis~\cite{AguilarSaavedra:2010sq,Azatov:2016sqh}. Furthermore, it
may occur that $\re (A^{(0)*} A^{(2)}_2)$ vanishes at well. This happens often
when the process is mediated by one heavy particle in the UV
theory~\cite{AguilarSaavedra:2011vw}, since its propagator generates effective
operators with the same symmetry properties at all orders. In this scenario, the
quadratic term $|A^{(1)}|^2$ and the interference term
$\re (A^{(0)*} A^{(2)}_{11})$ give the only corrections to order $1/\Lambda^4$
and the terms in $S_{2}$ are not necessary to compute the leading-order
correction to the cross section.

Is this situation preserved by field redefinitions in the effective theory? The
equivalence theorem tells us that the amplitudes are invariant and comparing
order by order we see that the same will hold for each $A^{(i)}$. However, the
individual contributions $A^{(2)}_2$ and $A^{(2)}_{11}$ need not be invariant
separately. Hence, it is possible that $\re (A^{(0)*} A_2^{\prime(2)})$ does not
vanish any longer, and then the new operators in $S^{\prime}_2$ cannot be neglected, unless they do not interfere with $A^{(1)}$.

The quadratic terms may also be very relevant if the coefficient of an involved
operator $\Ocal^{(1)}$ in $S_1$ is for some reason $\alpha > 1$. Then,
$|A^{(1)}|^2$ and $A^{(0)*} A^{(2)}_{11}$ are enhanced by $\alpha$ with respect to $A^{(0)*}
A^{(1)}$. All these terms could then be comparable at sufficiently high energies. In
this case, it is mandatory to include them. Furthermore, at second order the effect of operators in $S_2$
can be neglected if it is known that their coefficients are significantly
smaller than $\alpha^2$. This is the case in certain SM extensions (such as the
example in section~\ref{sec:example}). But again, these statements depend on the
field coordinates. A field redefinition that removes $\Ocal^{(1)}$
introduces in $S_2$ operators with an enhancement $\alpha^2$, so their
contributions $\re (A^{(0)*}A^{\prime(2)}_2)$ can no longer be neglected.

\subsection{The loop expansion}
\label{sec:loop-expansion}

Our previous discussion of power counting also applies to the loop expansion of the effective theory and of the fundamental theory. Let us start with the former, which makes no reference to loops in the fundamental theory and is valid also for strongly coupled UV theories.\footnote{In some interesting cases, the former are related with some other parameter in the fundamental theory. For instance, loops in chiral perturbation theory are related to $1/N_c$ corrections in low-energy QCD. This type of relation has been made precise in gauge-gravity dualities~\cite{Maldacena:1997re}.} Reintroducing explicitly $\hbar$, we can formally expand the generating functional of the renormalized effective theory as
\beq
Z^R[J] = \sum_{k=0}^\infty \hbar^k Z_{\text{eff}}^{R(k)}[J] , \label{eq:IR-loop-expansion}
\eeq
This actually corresponds to an expansion in the effective-theory couplings divided by $1/(4\pi)^2$. We have already mentioned that the power counting of the (renormalized) effective action should be consistent  with this expansion. 

When working in a reduced basis at order $\lambda^n$, it is often found that counterterms made out of operators that were removed to reach  that basis are necessary to obtain renormalized Green functions. These counterterms (including their arbitrary finite part) can then be written in the reduced basis, to order $n$, by a perturbative field redefinition in which the perturbation parameter is proportional to $\hbar^m \lambda^n$, with $m$ the loop order of the counterterm.  In this way, one finds a  reduced renormalized action $(S^R)^\prime$ (instead of the initial renormalized reduced action). As stressed in section~\ref{sec:reparametrization-invariance}, this action does not give finite Green functions of the elementary field when the regulator is removed. But it does give finite S-matrix elements. So, we can say that the theory described by this action has been renormalized on shell (this concept is not to be confused with an on-shell renormalization scheme). To illustrate this, consider one of the simplest examples of a reduced action: requiring canonical normalization of the kinetic terms in order to remove the exact ambiguity of field rescalings. To obtain finite Green functions, wave function renormalization is required. Then, the renormalized action is no longer in the reduced form. By a regulator-dependent field rescaling, we can, however, transform the renormalized action into a reduced renormalized action, which has canonical kinetic terms; the wave function counterterms are moved into a redefinition of the remaining counterterms. But the Green functions associated to this action are just the Green functions of the bare field (written in terms of renormalized masses and couplings), which are divergent~\cite{Collins:1984xc}. Nevertheless, these Green functions can be used to calculate finite scattering amplitudes, with the regulator removed after the on-shell reduction. 
Coming back to the general case, note that, at higher orders in $\lambda$, the reduced renormalized action will contain also corrections of order $\hbar^m$ and higher, as indicated in the power-counting formula eq.~\refeq{eq:power-counting}. These higher-order counterterms are also required for finiteness of the S matrix. 

Importantly, the finite parts of all the redefined counterterms can be fixed in terms of renormalization conditions for each operator in the reduced action~(see ref.~\cite{Ball:1993zy} for a detailed argument in the context of the exact renormalization group). Thus, no independent renormalized couplings associated to redundant operators need to be introduced. This implies that one can describe the renormalization-group evolution of the reduced renormalized couplings in terms of reduced renormalized couplings only, which has led to the definition in~\cite{Einhorn:2001kj} of effective beta functions along the reduced directions, depending only on reduced renormalized couplings. The renormalization group equation of on-shell quantities can be written in terms of these effective beta functions. Depending on the aimed precision, the higher-order corrections introduced by the field redefinition may be relevant for the running of reduced couplings. Once again, we stress that using the equations of motion may lead to incorrect results. 

The linearized renormalization-group evolution can be described in terms of operator mixing; in this case, the beta functions are just anomalous dimensions. It has been observed in theories of interest that, at one-loop, the anomalous-dimension matrix has many vanishing entries, not explained by power counting~\cite{Grojean:2013kd,Elias-Miro:2013gya,Elias-Miro:2013mua,Jenkins:2013zja,Jenkins:2013wua,Elias-Miro:2013eta,Alonso:2014rga}. This pattern has been explained in terms of the Lorentz structure of the involved operators, which forbids certain mixings at one loop~\cite{Elias-Miro:2014eia}.

Let us next consider the loop expansion of the fundamental theory $S_{\mathrm{UV}}$, which we assume to be weakly coupled:
\beq
Z^R[J] = \sum_{k=0}^\infty \hbar^k Z^{R(k)}[J] . \label{eq:UV-loop-expansion}
\eeq
This corresponds to an expansion in the UV couplings divided by $1/(4\pi)^2$. In order to match this expansion, the bare effective action $\bar{S}$ in eq.~\refeq{eq:matching} and its local version $S=\langle \bar{S} \rangle_n$ must depend explicitly on $\hbar$. 
Each coefficient $Z^{R(k)}$ is recovered by combining the powers of $\hbar$ in $S$ with the ones associated to loops (and counterterms) in calculations performed within the effective theory. 
Consider a double expansion of $S$ in $\hbar$ and $\lambda$:
\beq
S[\phi] = \sum_{m=0}^n \sum_{k=0}^\infty \hbar^k \lambda^m S_m^{(k)}[\phi].
\eeq
Note that if all the possible operators are included, then all the $S_m^{(k)}$ with a fixed $m$ will contain the same operators. That is, the corrections from matching at the quantum level can be absorbed into a renormalization of the coefficients. But as discussed above, the point of matching is to compare the renormalized parameters of the effective theory with the UV parameters in a renormalization scheme that is independent of the fundamental theory. Let us now perform a perturbative field redefinition to eliminate an operator in $S_n^{(j)}$. This will rearrange all $S_m^{(k)}$ with $k\geq j$ and $m\geq n$, in a way consistent with eq.~\refeq{eq:power-counting}. Once again, there are practical consequences for the matching workflow. Suppose, for example, that $S_0^{(1)}$ is non vanishing and that we want to eliminate a first order term at the classical level, that is, a term in $S_1^{(0)}$. Then, there will be corrections not only to $S_1^{(0)}$ but also to $S_1^{(1)}$. This means that to calculate the matching at one-loop one must not only integrate out at that level, but also keep track of possible rearrangements of the effective action at the classical level. For this, it is not sufficient to know the final form at the classical level, $(S^\prime)_1^{(0)}$. So, the necessary corrections would be missed if one simply added the one-loop result to the results of tree-level  matching given in the literature in particular basis. In other words and with more generality, the same light fields should be used in calculating the contributions at each order in the loop expansion.

A related issue is the fact that the classification in~\cite{Arzt:1994gp} of tree-level and loop operators, as those that can be induced or not at the tree-level, respectively, is not stable under field redefinitions. Therefore, this classification is only meaningful in one the following two interpretations: either for classes of operators that can be connected by field redefinitions, as proposed in~\cite{Elias-Miro:2013gya}, or for individual operators in the context of a given non-redundant basis of operators. This latter classification is basis-dependent. It turns out that the former is closely related to the pattern of operator mixing~\cite{Elias-Miro:2014eia}.

\section{Realistic example}
\label{sec:example}

We present here a realistic example at the tree level in the SMEFT, to illustrate some of the higher-order effects of field redefinitions. We start with an extension of the SM consisting
of a heavy hypercharge-neutral vector triplet $\mathcal{W}$ (see ref.~\cite{deBlas:2012qp,Pappadopulo:2014qza,Aaboud:2018bun} for the phenomenology of this multiplet), coupling only to
the Higgs doublet $\phi$, for simplicity. 
\begin{align}
  \Lcal
  &=
  \Lcal_{\text{SM}}
  + \frac{1}{2}
  \left(
    D_\mu \mathcal{W}^a_\nu D^\nu \mathcal{W}^{a\mu}
    - D_\mu \mathcal{W}^a_\nu D^\mu \mathcal{W}^{a\nu}
    - M^2 \mathcal{W}^a_\mu \mathcal{W}^{a\mu}
  \right)
  \nn
  &\phantom{=}
  - g \, \mathcal{W}^a_\mu
  \left(
    \phi^\dagger \frac{\sigma^a}{2} i D_\mu \phi
    + \hc
  \right).
\end{align}
Here, $g$ is a real coupling constant and $\Lcal_{\text{SM}}$ is the
SM Lagrangian:
\begin{align}
  \Lcal_{\text{SM}}
  &=
  D_\mu \phi^\dagger D^\mu \phi
  + \mu^2 \phi^\dagger \phi
  - \lambda (\phi^\dagger \phi)^2
  + \Lcal_{\text{Yuk.}} + \dots,
  \\
  \Lcal_{\text{Yuk.}}
  &=
  - \bar{l}_L Y_e \phi e_R
  - \bar{q}_L Y_d \phi d_R
  - \bar{q}_L Y_u \tilde{\phi} u_R
    + \hc
\end{align}

The tree-level matching to the SMEFT to dimension 8,
\begin{equation}
  \Lcal_{\text{eff}}
  =
  \Lcal_{\text{SM}}
  + \Lcal^{(6)}_{\text{eff}}
  + \Lcal^{(8)}_{\text{eff}}
  + O\left( \frac{1}{M^6} \right),
\end{equation}
gives
\begin{align}
  \Lcal^{(6)}_{\text{eff}}
  &=
  - \frac{3 g^2}{8 M^2} (\phi^\dagger \phi) \square (\phi^\dagger \phi)
  + \frac{g^2}{4 M^2}
  \left[
    (\phi^\dagger \phi) (\phi^\dagger D^2 \phi)
    + \hc
  \right],
  \\
  \Lcal^{(8)}_{\text{eff}}
  &=
  \frac{g^2}{8 M^4}
  D_\mu (\phi^\dagger \sigma^a i D_\nu \phi + \hc)
  \left[
    D^\nu (\phi^\dagger \sigma^a i D^\mu \phi)
    - D^\mu (\phi^\dagger \sigma^a i D^\nu \phi)
    + \hc
  \right],
\end{align}
where we have used the identity
$ \sigma^a_{ij} \sigma^a_{kl} = 2 \delta_{il} \delta_{jk} - \delta_{ij}
\delta_{kl} $ and integration by parts to simplify
$\Lcal^{(6)}_{\text{eff}}$. We see that the characteristic scale $\Lambda$ that controls the size of each operator is in this case equal to the mass $M$. Each operator in $\Lcal_{\text{eff}} - \Lcal_{\text{SM}}$ of canonical dimension $\Delta$ is proportional to $M^{4-\Delta}$. The expansion of the effective Lagrangian is in powers of $1/M^2$, not in powers of $g^2/M^2$; indeed, all the terms are proportional to $g^2$, independently of the dimension. This pattern is characteristic of integrating heavy particles out at the tree level.
 
Now, to remove the operator containing
$D^2 \phi$, we redefine the Higgs doublet:
\begin{equation}
  \label{eq:smeft-example-redefinition}
  \phi \to \phi + \frac{g^2}{4 M^2} \phi (\phi^\dagger \phi).
\end{equation}
The resulting Lagrangian is:
\begin{align}
  \Lcal^{\prime}_{\text{eff}}
  &=
  \Lcal_{\text{SM}}
  + \frac{\mu^2 g^2}{2 M^2} {(\phi^\dagger \phi)}^2
  - \frac{\lambda g^2}{M^2} {(\phi^\dagger \phi)}^3
  + \frac{g^2}{4 M^2} {(\phi^\dagger \phi)}^2 \Lcal_{\text{Yuk.}} 
  \label{eq:SM-redef-1}
  \nn
  &\phantom{=}
  + \left\{
    \frac{g^4}{16 M^4}
    D_\mu\left( (\phi^\dagger \phi) \phi^\dagger \right)
    D^\mu\left( \phi (\phi^\dagger \phi) \right)
    + \frac{\mu^2 g^4}{16 M^4} {(\phi^\dagger \phi)}^3
    - \frac{3 \lambda g^4}{8 M^4} {(\phi^\dagger \phi)}^4
  \right\}
  \nn
  &\phantom{=}
  - \frac{3 g^2}{8 M^2} (\phi^\dagger \phi) \square (\phi^\dagger \phi)
  - \frac{3 g^4}{8 M^4} {(\phi^\dagger \phi)}^2 \square (\phi^\dagger \phi)
  \nn
  &\phantom{=}
  + \frac{3 g^4}{16 M^4} {(\phi^\dagger \phi)}^2 (\phi^\dagger D^2 \phi + \hc)
  - \frac{g^4}{8 M^4}
  D_\mu\left( (\phi^\dagger \phi) \phi^\dagger \right)
  D^\mu\left( \phi (\phi^\dagger \phi) \right)
  \nn
  &\phantom{=}
  + \frac{g^2}{8 M^4}
  D_\mu (\phi^\dagger \sigma^a i D_\nu \phi + \hc)
  \left[
    D^\nu (\phi^\dagger \sigma^a i D^\mu \phi)
    - D^\mu (\phi^\dagger \sigma^a i D^\nu \phi)
    + \hc
  \right]
  \nn
  &\phantom{=}
  + O\left(\frac{1}{M^6}\right).
\end{align}
We have grouped the terms by their origin, rather than by the powers of $M^{-2}$. The terms in the first two lines of this equation arise from the
redefinition of $\Lcal_{\text{SM}}$, the ones
in the third and fourth lines come from
$\Lcal^{(6)}_{\text{eff}}$ and the fifth line is just
$\Lcal^{(8)}_{\text{eff}}$. The $O(1/M^4)$ contributions from
$\Lcal_{\text{SM}}$ (the terms inside curly brackets in the second line) are quadratic in the
perturbation of $\phi$ introduced by the redefinition.

We observe that removing the dimension-6 term proportional to $D^2\phi$ complicates the Lagrangian significantly in several ways. First, there are more operators. Second, the fact that the different operators are generated by the exchange of a vector triplet is far from apparent. Third, the physical implications are also more obscure, due to the essential correlations that guarantee that  the redefined Lagrangian $\Lcal^{\prime}_{\text{eff}}$ gives the same S matrix as $\Lcal_{\text{eff}}$ (and as $\Lcal$, at low energies). And finally, the simple features we had observed in $\Lcal_{\text{eff}}$ are not preserved: i) The powers of $M^{-2}$ are not determined by the canonical dimension of the operator:  the second terms of the first and second lines have dimension 4 and 6, respectively, but have an extra $\mu^2/M^2$ suppression. ii) Terms proportional to $g^4$ appear. Both issues have been discussed in subsection~\ref{subsec:powercounting}. 

It is also worth noting that some care must be taken when using these results
for matching at one loop, as pointed out in
section~\ref{sec:loop-expansion}. So far, we have been manipulating the
tree-level matching result $\Lcal_{\text{eff}}$. Let
$\Lcal^{\text{1-loop}}_{\text{eff}} = \Lcal_{\text{eff}} + \Delta
\Lcal_{\text{eff}}$ be the result of one-loop matching. It is clear that, from
$\Lcal^{\prime}_{\text{eff}}$ and $\Delta \Lcal_{\text{eff}}$ alone,
$\Lcal^{\text{1-loop}}_{\text{eff}}$ cannot be computed. It is necessary to
know the redefinition~\refeq{eq:smeft-example-redefinition} that has been done
in the tree-level matching results. Then, it could be applied in
$\Delta \Lcal_{\text{eff}}$ to get $\Delta \Lcal^\prime_{\text{eff}}$. This
could be used to obtain
${(\Lcal^{\text{1-loop}}_{\text{eff}})}^\prime = \Lcal^\prime_{\text{eff}} +
\Delta \Lcal^\prime_{\text{eff}}$, which is equivalent to
$\Lcal^{\text{1-loop}}_{\text{eff}}$.

We compare now $\Lcal^\prime_{\text{eff}}$ with the result of using the
higher-order equation of motion in $\Lcal_{\text{eff}}$. This exemplifies what
is discussed in section~\ref{sec:eom} and subsection~\ref{subsec:nonredundant}. The equation of motion of
$\Lcal_{\text{eff}}$ to dimension 6 is
\begin{align}
  D^2 \phi
  &=
  \mu^2 \phi
  - 2 \lambda \phi (\phi^\dagger \phi)
  + \frac{\delta \Lcal_{\text{Yuk.}}}{\delta \phi^\dagger}
  - \frac{3 g^2}{4 M^2} \phi \square (\phi^\dagger \phi)
  \nonumber
  \\
  &\phantom{=}
  + \frac{g^2}{4 M^2}
  \left[
    \phi (\phi^\dagger D^2 \phi)
    + D^2 \phi (\phi^\dagger \phi)
    + \phi (D^2 \phi^\dagger \phi)
    + D^2 \left( \phi (\phi^\dagger \phi) \right)
  \right]
  + O\left(\frac{1}{M^4}\right).
\end{align}
If we use it to eliminate the terms proportional to $D^2\phi$ and
$D^2 \phi^\dagger$ in $\Lcal_{\text{eff}}$, we get:
\begin{align}
  \Lcal^{\text{linear}}_{\text{eff}}
  &=
  \Lcal_{\text{SM}}
  + \frac{\mu^2 g^2}{2 M^2} {(\phi^\dagger \phi)}^2
  - \frac{\lambda g^2}{M^2} {(\phi^\dagger \phi)}^3
  + \frac{g^2}{4 M^2} (\phi^\dagger \phi) \Lcal_{\text{Yuk.}} 
  \nn
  &\phantom{=}
  - \frac{3 g^2}{8 M^2} (\phi^\dagger \phi) \square (\phi^\dagger \phi)
  - \frac{3 g^4}{8 M^4} {(\phi^\dagger \phi)}^2 \square (\phi^\dagger \phi)
  \nn
  &\phantom{=}
  + \frac{3 g^4}{16 M^4} {(\phi^\dagger \phi)}^2 (\phi^\dagger D^2 \phi + \hc)
  - \frac{g^4}{8 M^4}
  D_\mu\left( (\phi^\dagger \phi) \phi^\dagger \right)
  D^\mu\left( \phi (\phi^\dagger \phi) \right)
  \nn
  &\phantom{=}
  + \frac{g^2}{8 M^4}
  D_\mu (\phi^\dagger \sigma^a i D_\nu \phi + \hc)
  \left[
    D^\nu (\phi^\dagger \sigma^a i D^\mu \phi)
    - D^\mu (\phi^\dagger \sigma^a i D^\nu \phi)
    + \hc
  \right]
  \nn
  &\phantom{=}
  + O\left(\frac{1}{M^6}\right).
\end{align}
We see that the quadratic terms in the second line of eq.~\refeq{eq:SM-redef-1}, in curly brackets, are absent, as expected.\footnote{A wrong result is found as well when imposing the dimension-6 equations of motion of $\Lcal-\frac{g^2}{4 M^2}
  \left[
    (\phi^\dagger \phi) (\phi^\dagger D^2 \phi)
    + \hc
  \right]$ or, iteratively, of $\Lcal^\prime_{\text{eff}}$.} In particular, the dimension-6 term proportional to $\mu^2$ is missing. 
These terms can be recovered by including the second derivative of the action, $1/2  \delta^2 S_{\text{eff}}/{(\delta \phi^\alpha \delta \phi^\beta)}
  f^\alpha f^\beta$, with $f^1 = g^2/(4M^2)  \phi (\phi^\dagger \phi)$ and $f^2 = (f^1)^\dagger$. We find
\begin{align}
  \Lcal^{\text{quad.}}_{\text{eff}}
  &=
  \frac{\mu^2 g^4}{16 M^4}
  (\phi^\dagger \phi)^3
  + \frac{g^4}{16 M^4}
  D_\mu \left( (\phi^\dagger \phi) \phi^\dagger \right)
  D^\mu \left( (\phi^\dagger \phi) \phi \right)
  - \frac{3 \lambda g^4}{8 M^4}
  (\phi^\dagger \phi)^4
  + O\left( \frac{1}{M^6} \right).
\end{align}
It is clear that
$
\Lcal^\prime_{\text{eff}}
=
\Lcal^{\text{EOM}}_{\text{eff}}
+ \Lcal^{\text{quad.}}_{\text{eff}}
+ O(1/M^6)
$.

The equivalence between $\Lcal^\prime_{\text{eff}}$ and $\Lcal_{\text{eff}}$ can
be also checked using the condition for redundant parameters stated in
section~\ref{sec:eom}. We define the interpolating Lagrangian:
\begin{equation}
  \Lcal^\xi_{\text{eff}}[\phi]
  =
  \Lcal_{\text{eff}}
  \left[
    \phi + \frac{\xi}{4 M^2} \phi (\phi^\dagger \phi)
  \right],
\end{equation}
which is just $\Lcal_{\text{eff}}$ when $\xi = 0$ and
$\Lcal^\prime_{\text{eff}}$ when $\xi = g^2$. Then, $\Lcal_{\text{eff}}$ and
$\Lcal^\prime_{\text{eff}}$ are equivalent because $\xi$ is redundant:
\begin{equation}
  \frac{\partial \Lcal^\xi_{\text{eff}}}{\partial \xi}
  =
  \left(
    \frac{1}{4 M^2} (\phi^\dagger \phi) \phi^\dagger
    - \frac{3 \xi}{16 M^4} (\phi^\dagger \phi)^2 \phi^\dagger
  \right)
  \frac{\delta \Lcal^\xi_{\text{eff}}}{\delta \phi^\dagger}
  + \hc
  + O\left(\frac{1}{M^6}\right).
\end{equation}

\section{Conclusions}
\label{sec:conclusions}

It is clear that a perturbative transformation, controlled by a small parameter $\lambda$, of any function depending analytically on $\lambda$ will rearrange at all orders its perturbative expansion in $\lambda$, with the new coefficients depending on the original ones of the same or lower order. It is also clear that this rearrangement cannot be reproduced by a linear approximation in the perturbation. These simple facts may have non-trivial practical implications for effective field theories.

Effective field theories are treated perturbatively in $1/\Lambda$
and in a loop expansion. When putting together different orders, it is crucial
that they are all given in the same field coordinates. Otherwise,
inconsistencies will be present, not only off-shell but also in on-shell
observables. Preserving the consistency of field redefinitions requires some
care when the different orders are calculated independently. Consider, for
example, the SMEFT. A complete matching of this effective theory to arbitrary UV
completions has been given in~\cite{deBlas:2017xtg} at the tree-level and to
order $1/\Lambda^2$, with $\Lambda$ the lightest mass of the heavy
particles. The results of the matching are given in the Warsaw
basis~\cite{Grzadkowski:2010es}. They are very useful when working to order
$1/\Lambda^2$ and at the tree level, but, unfortunately, they cannot be combined
with future direct results of tree-level matching at order $1/\Lambda^4$. For
this, knowledge of the higher-order terms generated by the lower-order field
redefinitions is required. But this information is usually not provided in the
literature, including~\cite{deBlas:2017xtg}, nor can it be recovered without
repeating the whole calculation. Similarly, the Warsaw-basis results of
tree-level matching cannot be combined with one-loop corrections, even if the
latter are transformed into the Warsaw basis. Moreover, in some methods it may
be convenient to also perform field redefinitions in the UV action in order to
find one-loop corrections to the matching. For consistency, the
tree-level contributions must be calculated for the same light fields. Note
that an identical situation will arise again and again at higher and higher
orders. This is not a fundamental problem, but it conflicts with the idea of
building on previous results. The very same issues are relevant for conversions
from one basis into another one. In particular, the generalization to higher orders of codes that automatically reduce actions~\cite{Criado:2017khh,Bakshi:2018ics} or translate operator coefficients in different
bases~\cite{Falkowski:2015wza,Gripaios:2018zrz} should implement field
redefinitions rather than use equations of motion.

Field redefinitions not only change the action, but they also introduce a
determinant (which can be added to the action or ignored in dimensional
regularization, for local perturbative redefinitions) and modify the coupling to
the sources. The latter effect is crucial in the derivation of Schwinger-Dyson
equations and Ward identities. Ignoring it amounts to the bold replacement of a
coupling of the source to a sum of composite operators by a linear coupling to
the new elementary field. The LSZ formula implies that this replacement has no
effect on on-shell quantities. But, as we have discussed, it does have a
non-trivial impact on the form of the local effective action after matching and
also on renormalization.
All these subtle effects are relevant for the standard approach to matching and renormalization in terms of Green functions. However, we should stress that they go away when computing on-shell amplitudes, and might be avoided from the beginning in on-shell matching/renormalization. 

Working with non-redundant bases of operators in effective theories has become a standard practice. Besides having a reduced number of operators, these bases have the clear advantage of attaching an unambiguous physical meaning to the set of coefficients that describe the theory to a given order. In particular, flat directions are avoided in comparing with the experimental data. Notwithstanding this, the conversion into non-redundant or reduced bases also has a few drawbacks. The first one is apparent in the example of section~\ref{sec:example} and, more dramatically, in the example of appendix~\ref{app:free}: the necessary field redefinitions typically give rise to a more complicated Lagrangian. Of course, this is not so in a truly model-independent approach, in which the starting point is a completely general effective theory. But even in this case, the connection to particular UV completions is more intricate. More importantly, the physical predictions are typically more obscure, as the redefinition introduces correlations between operator coefficients that must be precisely preserved.\footnote{This is also a consequence of some algebraic manipulations performed to reach a given basis, such as Fierz reorderings.} For instance, at first sight it is far from obvious that eq.~\refeq{eq:transformed-phi-lagrangian} represents a free theory in disguise. Another issue that we have discussed is that reduced actions are not stable under renormalization and renormalization-group evolution, although the departures can be absorbed on-shell into reduced counterterms and effective beta functions. Finally, we have seen that field redefinitions may modify the power counting inherited from (classes of) UV theories, when it cannot be formulated in terms of the effective theory alone. So, such a power counting needs not be apparent in non-redundant bases. 

The basis proposed in~\cite{Elias-Miro:2013gya,Elias-Miro:2013mua} is optimal in dealing with all the issues just mentioned, but only for particular processes (Higgs physics) and rather specific UV scenarios (universal theories). 
Let us put forward another possibility: working with a standard over-complete, i.e.\ non-reduced, basis. In principle, this minimizes the problems pointed out above. Indeed, the connection with UV theories is more transparent, at least at the tree level, and there is flexibility in reproducing the field coordinates used in the matching. Also, if no redefinitions are made after matching, the physical predictions will typically be more obvious, and for simple models will not contain flat directions. The tree-level or loop origin of operators is directly given by the classification in~\cite{Arzt:1994gp}. And finally, from the point of view of the effective theory itself, a general action in the over-complete basis is stable under renormalization and gives rise to finite off-shell Green functions that obey standard renormalization group equations.  Working in this approach would first involve selecting a basis at each order, obtained only with algebraic manipulations of the operators (the convenience of the latter should also be assessed in each case). Then, the results of matching and the beta functions would be provided in this basis (with information about possible field redefinitions in the process). And finally, to profit from the advantages of reduced bases, it would be useful to know the conversions of the over-complete basis into non-redundant bases, including higher-order operators generated in the process, or to have the tools to perform automatically this task.

\acknowledgments

We thank Juan Antonio Aguilar Saavedra and Jos\'e Santiago for useful
discussions. Our work has been supported by the Spanish MINECO project
FPA2016-78220-C3-1-P (Fondos FEDER) and the Junta de Andaluc\'ia grant
FQM101. The work of J.C.C.\ has also been supported by the Spanish MECD grant
FPU14.

\appendix

\section{Simple example of reparametrization invariance}
\label{app:free}

We describe here an example of a field redefinition in a simple quantum field
theory demonstrating explicitly some of the features explained in
section~\ref{sec:reparametrization-invariance}. We start with a free massless
real scalar field $\phi$. Its generating function is
\begin{equation}
  \label{eq:original-Z-example}
  Z[J]
  =
  \int D\phi
  \exp{
    \left(
      - \frac{i}{2} \phi_x (\square \phi)^x
      + J_x \phi^x
    \right)
  }.
\end{equation}

A change of variables $\phi \to \phi + (1/m^2) (\square \phi + g \phi^3)$ in the
path integral gives the following expression, where we have used equation
\refeq{eq:jacobian-ghosts}:
\begin{equation}
  \label{eq:transformed-Z-example}
  Z[J]
  =
  \int D\phi Dc D\bar{c}
  \exp{
    \left(
      i S[\phi, c, \bar{c}]
      + J_x \phi^x
      + \frac{1}{m^2} J_x \square \phi^x
      + \frac{g}{m^2} J_x (\phi^3)^x
    \right)
  },
\end{equation}
where $S[\phi, c, \bar{c}] = S_\phi[\phi] + S_c[\phi, c, \bar{c}]$
is given by
\begin{align}
  \label{eq:transformed-phi-lagrangian}
  S_\phi
  &=
  - \int d^d x \,
  \left[
    \frac{1}{2} \phi
    \,
    \square
    \left(
      1
      + \frac{\square}{m^2}
    \right)^2
    \phi
    + \frac{g}{m^2} \phi^3
    \square
    \left(
      1
      + \frac{\square}{m^2}
    \right)
    \phi
    + \frac{g^2}{2 m^4} \phi^3 \square \phi^3
  \right],
  \\
  \label{eq:ghost-lagrangian}
  S_c
  &=
  - \int d^d x \,
  \left[
    \bar{c} (\square + m^2) c
    + 3 g \phi^2 \bar{c} c
  \right].
\end{align}
We have normalized $c$ to have a canonical kinetic term. The momentum-space
connected Green functions are defined as
\begin{equation}
\label{eq:Fourier}
  G^{(n)}(p_1, \ldots, p_n)
  =
  a_{x_1}(p_1) \cdots a_{x_n}(p_n)
  \frac{\delta^n W}{\delta J_{x_1} \cdots \delta J_{x_n}},
\end{equation}
where $a_{x}(p) = e^{ipx}$ and $W[J] =-i\log{Z[J]}$. Let us show diagrammatically
that using expression eq.~\refeq{eq:transformed-Z-example} for $Z$ to compute
$G^{(n)}$ gives the same result as using the original
form eq.~\refeq{eq:original-Z-example}.

\begin{figure}[t]
  \centering
  \includegraphics[width=14cm]{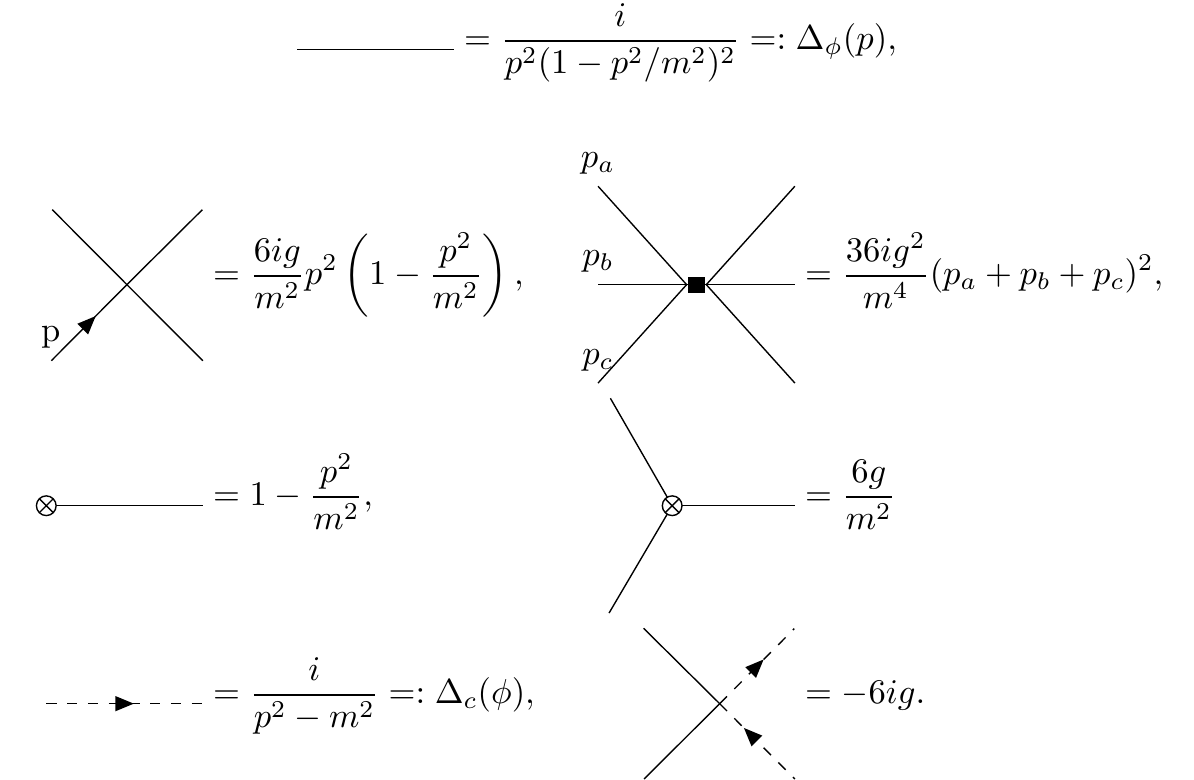}
  \caption{Feynman rules from eqs.~\refeq{eq:transformed-Z-example},
    \refeq{eq:transformed-phi-lagrangian}, \refeq{eq:ghost-lagrangian}. Crossed
    dots represent sources. Solid and dotted lines correspond to $\phi$ and
    ghosts, respectively. An arrow over a $\phi$ line is used to specify that
    the corresponding momentum enters in the factor associated with the vertex
    it points to. The square that splits the 6-line vertex specifies the three
    momenta that appear in its associated factor.}
  \label{fig:feynman-rules}
\end{figure}

$G^{(n)}$ is the sum over all connected diagrams with $n$ sources constructed
using the Feynman rules collected in Figure~\ref{fig:feynman-rules}.  The
propagator $\Delta_\phi(p)$ for $\phi$ contains the physical pole at $p^2 = 0$
but also a new (double) pole at $p^2 = m^2$ that was not present
originally. This problematic behavior will be canceled by the momentum
dependent vertices and the pole at the same point of the ghost propagator
$\Delta_c(p)$.

\begin{figure}[t]
  \centering
  \includegraphics[width=14cm]{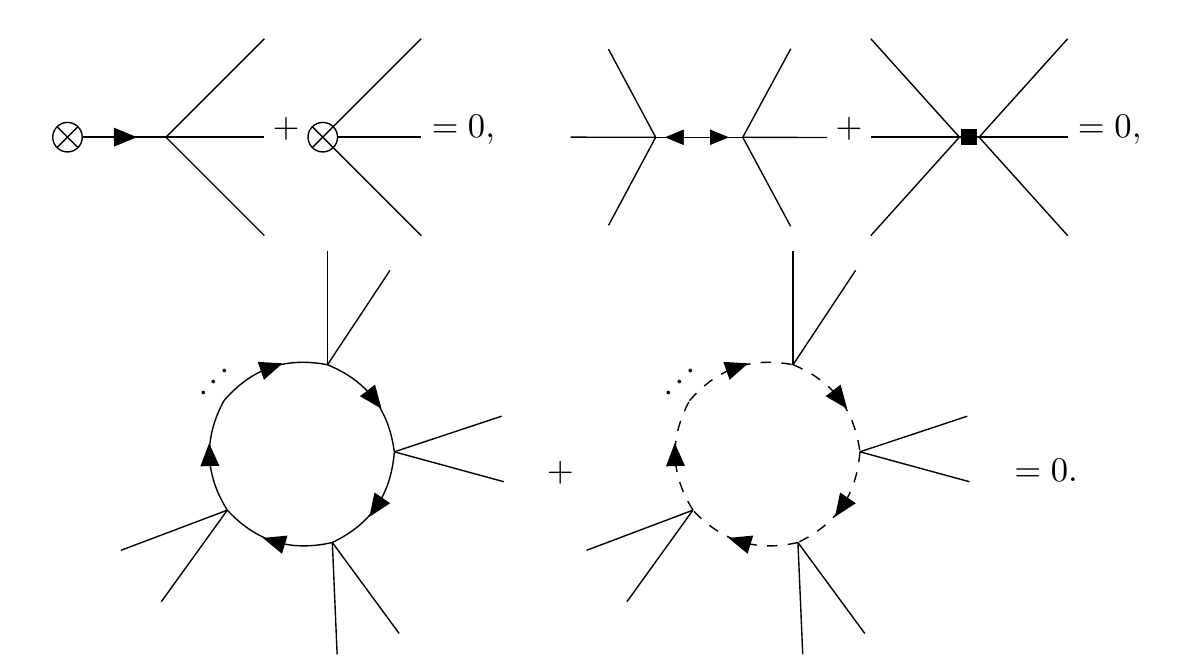}
  \caption{Cancellations between subdiagrams.}
  \label{fig:cancellations}
\end{figure}

There are several cancellations between subgraphs of the diagrams we are
considering. This is just an example of the more general case nicely discussed
in~\cite{tHooft:1973wag}. Three of these cancelllations are shown in
Figure~\ref{fig:cancellations}. From the first two equations in this figure, it
follows that that we can obtain the full result by summing over a subset of all
diagrams: those that do not contain 3-line sources, 6-line vertices, any arrows
in external lines or two arrows in the same internal line.  In other words, we
only need to consider diagrams with 1-line sources, 4-line vertices, no arrows
in external lines and at most one arrow in each internal line.

For any diagram, let $V$ be the number of vertices, $I$ the number of internal
lines and $L$ the number of loops. We have the relation
\begin{equation}
  \label{eq:eulers-formula}
  V - I + L = 1.
\end{equation}
The number of arrows over $\phi$ lines equals the number of $\phi^4$ vertices,
so at tree level ($L = 0$) there are no diagrams with less than two arrows in
all internal lines. The only exception is the case $V = 0$, which gives the only
diagram contributing to $G^{(2)}(p, -p) = i/p^2$. All the other Green functions
vanish at tree level.

For $L > 0$, we can reduce the problem by cutting all internal $\phi$ lines
without arrows. The result might be disconnected. For each connected component
$C$ the number of $\phi^4$ vertices equals the number of internal $\phi$ lines
and the number of $\phi^2 \bar{c}c$ vertices equals the number of ghost
lines. Therefore, using  eq.~\refeq{eq:eulers-formula}, $C$ has exactly one
loop ($L = 1$). A 1-loop diagram has as a subgraph one of the two 1-loop
diagrams in Figure~\ref{fig:cancellations}, so it must cancel with the diagram
obtained by replacing the subgraph with the other 1-loop diagram in the same
figure. The cancellation of the connected components after the cut implies the
cancellation of the diagrams resulting from joining them back. The conclusion is
that the $L$-loop correction (with $L > 0$) to any Green function is zero.

We have computed all the Green functions to all orders in the loop expansion:
\begin{equation}
  G^{(2)}(p, -p) = \frac{i}{p^2}, \qquad G^{(n > 2)} \equiv 0,
\end{equation}
using $Z$ in the form of  eq.~\refeq{eq:transformed-Z-example}. They agree
exactly with what is obtained in a more straightforward way from the expression
in eq.~\refeq{eq:original-Z-example}. Therefore, they must also be equal
order by order in $p^2 / m^2$. This means that if we had worked perturbatively
in $p^2 / m^2$ we would have obtained the same results. As stated in general in
section~\ref{sec:reparametrization-invariance}, we can ignore the ghosts when
the redefinition is perturbative and dimensional regularization is used. Indeed, the ghost propagator should then be expanded in $1/m^2$ up to some finite order $k$ as
\beq
\Delta_c(p)  \mapsto \frac{-i}{m^2} \left(1+ \frac{p^2}{m^2}+\frac{p^4}{m^4} + \dots + \frac{p^{2k}}{m^{2k}} \right).
\eeq
This is a polynomial in $p^2$, so it integrates to zero in the ghost loop integral over $p$, according to the properties of dimensional regularization. 
In the same manner, in a perturbative treatment, the propagator of the redefined field $\phi$ is to be expanded in $1/m^2$ as
\beq
\Delta_\phi(p)  \mapsto 
\frac{i}{p^2} + \frac{i}{m^2}  \left(2+ 3\frac{p^2}{m^2}+4\frac{p^4}{m^4} + \dots + (k+2) \frac{p^{2k}}{m^{2k}} \right).
\eeq
The first term in the second line is the original propagator, while the sum of the new terms proportional to $1/m^2$ is a polynomial in $p^2$. So, no unphysical poles appear perturbatively. Clearly,  ignoring the ghost contributions and using at the same time the exact (resummed) propagator $\Delta_\phi(p)$ would be inconsistent and would give rise to unphysical poles in the Green functions.

The Green functions $(G')^{(n)}$ generated with the function $Z'$, obtained from
 eq.~\refeq{eq:transformed-Z-example} by replacing
$J_\alpha F^\alpha(\phi) \to J_\alpha \phi^\alpha$, are equal to the ones
computed from $Z$ except for the source factors. Now, there is nothing to
cancel the first diagram in Figure~\ref{fig:feynman-rules}, but the
corresponding factor has a pole at $p^2 = m^2$ and not at $p^2 = 0$, so its
contribution is eliminated by the LSZ formula. The other difference, the $p^2 / m^2$
term in the factor corresponding to the 1-line source, also vanishes on shell.
Thus, $Z \sim Z'$.

The equivalence $Z \sim Z'$ can also be checked using the condition (stated in
section~\ref{sec:eom}) that the derivative of the action with respect to a
parameter is proportional to the equation of motion if and only if the parameter
is redundant. In this case,
\begin{equation}
    \frac{\partial S_\phi}{\partial (1/m^2)}
  =
  \frac{\delta S_\phi}{\delta \phi^x}
  \left(
    \square \phi
    + g \phi^3
    - \frac{1}{m^2}
    \left[
      \square^2 \phi 
      + 3 g \phi^2 \square \phi
      + g \square \phi^3
      + 3 g^2 \phi^5
    \right]
  \right)^x
  + O\left(\frac{1}{m^4}\right),
\end{equation}
which means that the parameter $1/m^2$ of the action $S_\phi$ is redundant. For
the parameter $g$, a similar equation (of the form
$\partial S_\phi/\partial g \propto \delta S_\phi /\delta \phi$) can be
obtained. However, this is not necessary to eliminate $g$ from $S_\phi$ because
$1/m^2$ can be taken to be zero (as it is redundant) and then $S_\phi$ becomes
independent of $g$.

\section{Insertions of the equation-of-motion operator in on-shell amplitudes}
\label{app:eom-counterexample}

\begin{figure}[t]
  \centering
  \includegraphics[width=14cm]{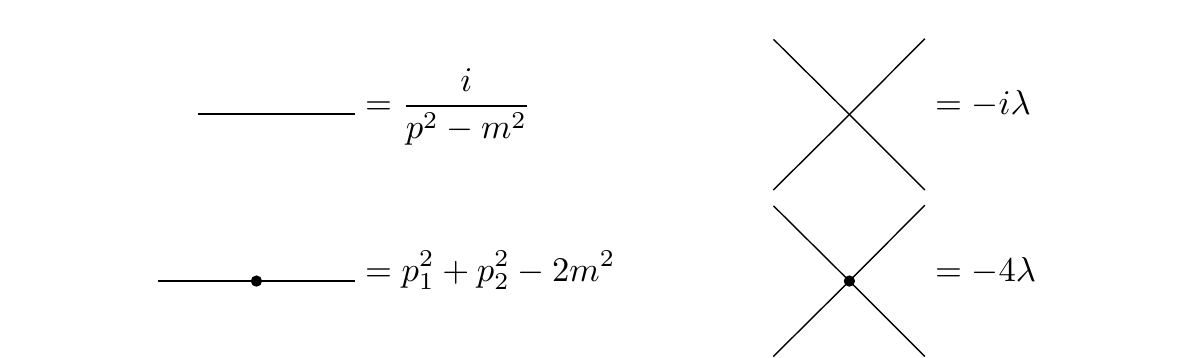}
  \caption{Feynman rules for $\phi^4$ theory and insertions of the operator
    $\theta$, represented by a solid dot.}
  \label{fig:eom-example-rules}
\end{figure}

\begin{figure}[t]
  \centering
  \includegraphics[width=12cm]{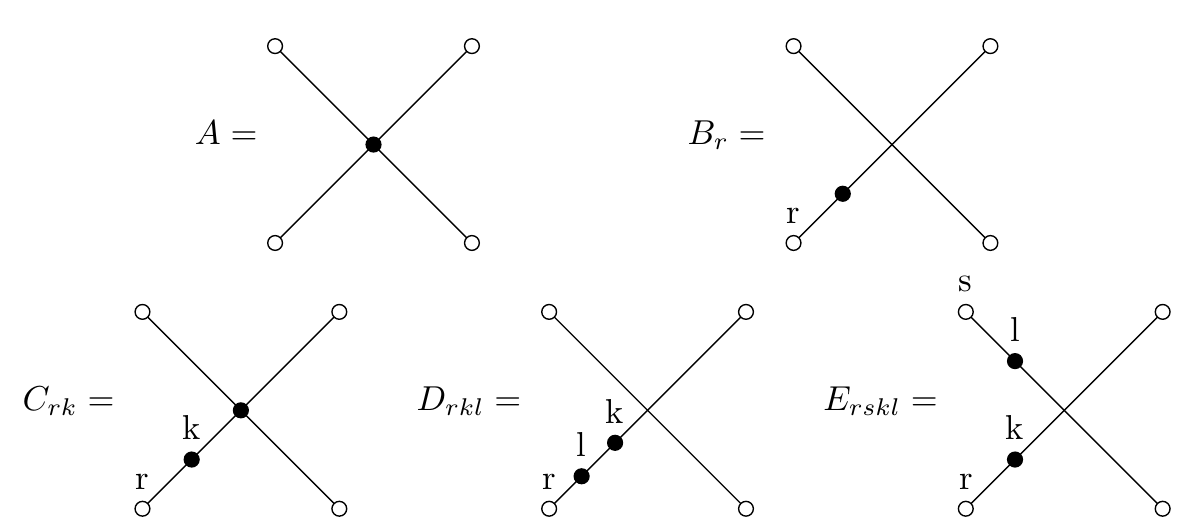}
  \caption{Relevant diagrams for the computation of $G^{(4, 1)}$ and
    $G^{(4, 2)}$ in the $\phi^4$ theory with $\theta$ insertions at tree
    level. Empty (solid) dots denote the sources for $\phi$ ($\theta$).}
  \label{fig:eom-example-diagrams}
\end{figure}

In ref.~\cite{Politzer:1980me} (see also ref.~\cite{Manohar:2018aog}) it is proven that the
S matrix with one insertion of an operator proportional to the equation of
motion vanishes. This is not true, however, for two or more insertions. We check
here both statements in the case proposed in exercise~6.1
of~\cite{Manohar:2018aog}. We will compute connected momentum-space Green
functions $G^{(m, n)}$ in the theory
\begin{align}
  Z[J^\phi, J^\theta]
  &=
  \int \Dcal \phi
  \exp{
    \left(
      i S[\phi]
      + J^\phi_x \phi^x
      + J^\theta_x \theta^x
    \right)
  }
  \\
  S[\phi]
  &=
  - \int d^4x
  \left(
    \frac{1}{2} \phi (\square + m^2) \phi
    + \frac{\lambda}{4!} \phi^4
  \right),
 \\
  \theta
  &=
  \phi \frac{\delta S}{\delta \phi}
  =
  - \phi (\square + m^2) \phi
  - \frac{\lambda}{3!} \phi^4.
\end{align}
They are defined in eq.~\refeq{eq:Fourier}.
The corresponding Feynman rules are presented in
Figure~\ref{fig:eom-example-rules}. We will calculate $G^{(4, 1)}$ and
$G^{(4, 2)}$. The relevant diagrams are shown in
Figure~\ref{fig:eom-example-diagrams}. In terms of them, the Green functions
are
\begin{align}
  \label{eq:green-1}
  G^{(4, 1)}
  &=
  A + \sum_{r = 1}^4 B_r,
  \\
  G^{(4, 2)}
  &=
  \sum_{r = 1}^4 \sum_{k = 1}^2 C_{rk}
  + \sum_{r = 1}^4
  \sum_{
    \substack{
      k,l = 1 \\
      k \neq l
    }
  }^2
  D_{rkl}
  + \sum_{
    \substack{
      r, s = 1 \\
      s > r
    }
  }^4
  \sum_{
    \substack{
      k, l = 1 \\
      k \neq l
    }
  }^2
  E_{rskl}.
\end{align}

The S matrix is obtained by taking the residue when all $p_i$ go on-shell.
Let $\res$ be the operation
\begin{equation}
  \res(G)
  =
  \lim_{p_1^2 \to m^2}
  \lim_{p_2^2 \to m^2}
  \lim_{p_3^2 \to m^2}
  \lim_{p_4^2 \to m^2}
  \left[
    \left(
      \prod_{i = 1}^4 (p_i^2 - m^2)
    \right)
    G
  \right].
\end{equation}
Applying it to each diagram gives
\begin{gather}
  \res(A) = - 4 \lambda,
  \qquad
  \res(B_r) = \lambda,
  \\
  \res(C_{rk}) = -4 i \lambda,
  \qquad
  \res(E_{rskl}) = i \lambda,
  \\
  \res(D_{rkl}) = i \lambda
  \left(
    1
    + \frac{(q_l + p_r)^2 - m^2}{(q_l + q_k + p_r)^2 - m^2}
  \right),
  \label{eq:diag-last}
\end{gather}
where all momenta are taken as ingoing. Using
equations~\eqref{eq:green-1}-\eqref{eq:diag-last} we get
\begin{align}
  \res(G^{(4, 1)})
  &=
  0,
  \\
  \res(G^{(4, 2)})
  &=
  i \lambda
  \left(
    -12
    + \sum_{r = 1}^4
    \sum_{
      \substack{
        k,l = 1 \\
        k \neq l
      }
    }^2
    \frac{(q_l + p_r)^2 - m^2}{(q_l + q_k + p_r)^2 - m^2}
  \right).
\end{align}
So, indeed, the S-matrix element with one insertion of $\theta$
vanishes. However, when two insertions of $\theta$ are included, it does not.

\section{Local vs non-local action after field redefinitions}
\label{app:simple-matching-counterexample}

We give here a counterexample to eq.~\refeq{eq:tilde-redefined-sources}.  This
illustrates how doing redefinitions does not commute with matching to a local
action. Notice that, instead of considering redefinitions of the fields in the
action, one can equivalently deal with redefinitions in the source terms,
because changes of variables in the path integral relate one case to the
other. We will use this fact to simplify the following discussion, in which we
consider changes of the source terms only.

\begin{figure}
  \centering
  \includegraphics[width=16cm]{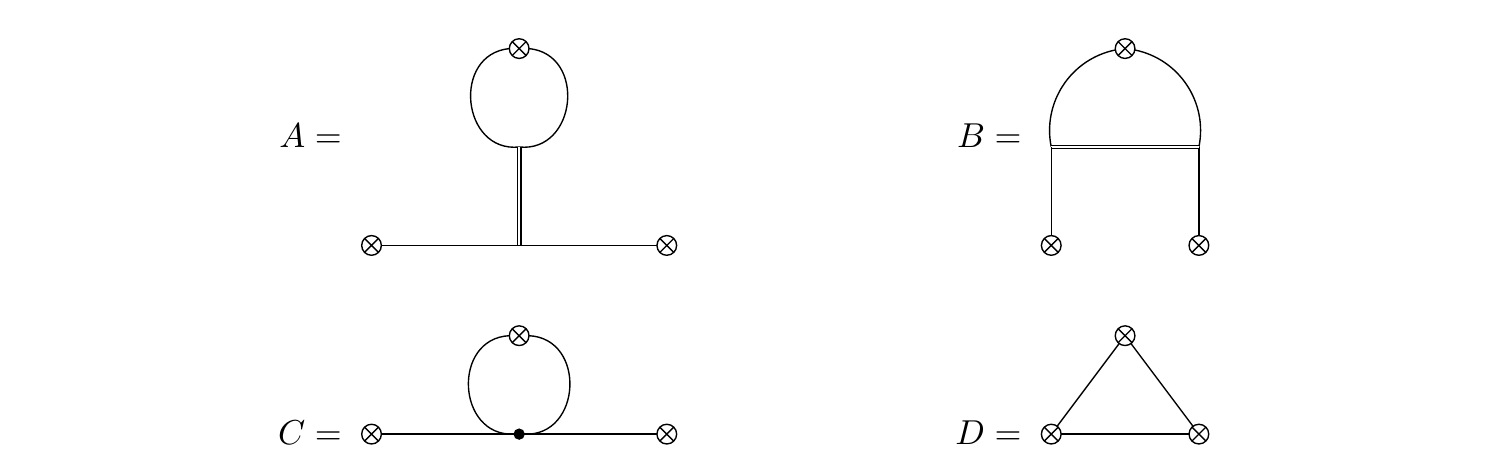}
  \caption{Relevant diagrams for the 1-loop 3-point function generated by
    $Z^\prime$ and $Z^\prime_n$. $A$ and $B$ are diagrams of $G^{\prime (3)}$,
    $C$ is a diagram of $G^{\prime (3)}_n$ and $D$ appears in both.
  }\label{fig:counterexample-diagrams}
\end{figure}

Consider the (non-local) action $\bar{S}$ coming from integrating out the field
$\Phi$, using  eq.~\refeq{eq:effaction}, from the theory defined by the UV
action
\begin{align}
  S_{\text{UV}}[\Phi, \phi]
  =
  -\int d^4 x
  \left\{
    \frac{1}{2} \phi \square \phi
    + \frac{1}{2} \Phi (\square + M^2) \Phi
    + g \Phi \phi^2
  \right\}.
\end{align}
Let $\bar{S}_{\text{tree}}$ be the action obtained by integrating out $\Phi$ at
tree-level. We take $1/M^2$ as the small parameter that controls the
approximation of the effective theory. The truncation of
$\bar{S}_{\text{tree}}$ is
\begin{equation}
  \truncated{\bar{S}_{\text{tree}}}{n}[\phi]
  =
  - \int d^4x
  \left\{
    \frac{1}{2} \phi \square \phi
    - \frac{g^2}{2 M^2}
    \phi^2
    \left(
      \sum_{k = 0}^{n - 1}
      \frac{{(-1)}^k \square^k}{M^{2k}}
    \right)
    \phi^2
  \right\},
\end{equation}
At tree-level, $\truncated{\bar{S}_{\text{tree}}}{n}$ gives the same results as
$\bar{S}_{\text{tree}}$ up to order $M^{-2n}$. The local effective action
$\corrected{\bar{S}}{n}$ is obtained by including both the heavy loop
corrections $\bar{S} - \bar{S}_{\text{tree}}$ and the corrections
$\corrected{\bar{S}}{n} - \truncated{\bar{S}}{n}$ due to heavy-light loops.
Notice that $\corrected{\bar{S}}{n}$ will not contain monomials that are odd
powers of $\phi$ because of the $\phi \to -\phi$ symmetry of the original action
$S_{\text{UV}}$, that is preserved in the effective theory. We will show that
the functions
\begin{align}
  Z^\prime[\phi]
  &=
  \int \Dcal\phi
  \exp{(i \bar{S}[\phi] + J_\alpha (\phi + \lambda \phi^2)^\alpha)},
  \\
  Z^\prime_n[\phi]
  &=
  \int \Dcal\phi
  \exp{(i \corrected{\bar{S}}{n}[\phi] + J_\alpha (\phi + \lambda \phi^2)^\alpha)},
\end{align}
do not satisfy the identity
$
Z^\prime[J]
\overset{?}{=}
Z^\prime_n[J]
+ O(1/M^{2n})
$
for any $n > 0$. It is enough to see that the 3-point functions $G^{\prime(3)}$ and
$G^{\prime(3)}_n$ generated by them are different. The relevant diagrams are
presented in Figure~\ref{fig:counterexample-diagrams}. Because computing Green
functions for $\phi$ with the non-local action $\bar{S}$ is exactly equivalent
to computing them with the local action $S_{\text{UV}}$, we present the
corresponding diagrams in terms of the Feynman rules for $S_{\text{UV}}$, with
double lines representing the propagator for the heavy field $\Phi$. The 4-line
dot in diagram $C$ represents the $\phi^4$ local interaction in
$\corrected{\bar{S}}{n}$ generated at tree level. We have
\begin{equation}
  G^{\prime (3)} = A + B + D + \text{(permutations)},
  \qquad
  G^{\prime (3)}_n = C + D + \text{(permutations)}.
\end{equation}

Diagram $C$ can be obtained by expanding in powers of $1/M^2$ the heavy
propagator inside $A + B$. Thus, for $G^{\prime (3)}_n$ to be equal to
$G^{\prime (3)}$ to order $n$, we should have
\begin{equation}
  A + B + (\text{permutations})
  \overset{?}{=}
  C + (\text{permutations})
  + O\left(\frac{1}{M^{2n + 2}}\right).
\end{equation}
This is not true in general. Denoting by $p_1$, $p_2$ and $p_3$ the momenta
in each diagram entering in the top, left and right vertex, respectively,
we have that, in dimensional regularization,
\begin{equation}
  A|_{p_1^2 = 0} = C|_{p_1^2 = 0} = 0,
\end{equation}
because when $p_1^2 = 0$ both $A$ and $C$ are scaleless integrals.
On the other hand,
\begin{align}
  B|_{p_1^2 = 0}
  &=
  -\frac{8 i g^2 \lambda \mu^{2 \epsilon}}{p_2^2 p_3^2}
  {
    \left.
      \int \frac{d^{4 - 2 \epsilon}k}{(2 \pi)^{4 - 2 \epsilon}}
      \frac{
        1
      }{
        k^2 (k + p_1)^2 [(k + p_1 + p_2)^2 - M^2]
      }
    \right|
  }_{p_1^2  = 0}
  \\
  & =
  \frac{g^2 \lambda}{2 \pi^2 p_2^2 p_3^2 (p_2^2 - p_3^2)}
  \Bigg\{
  \log
  \left(
    \frac{M^2 - p_3^2}{M^2 - p_2^2}
  \right)
  \left[
    \frac{1}{\epsilon}
    - \gamma
    + \log \frac{4 \pi \mu^2}{M^2}
  \right]
  + \operatorname{Li}_2 \frac{p_2^2}{M^2}
  - \operatorname{Li}_2 \frac{p_3^2}{M^2}
  \nonumber
  \\
  &\phantom{
    =
    \frac{g^2 \lambda}{2 \pi^2 p_2^2 p_3^2 (p_2^2 - p_3^2)}
    \Bigg\{
  }
  + \log^2 \left(1 - \frac{p_2^2}{M^2}\right)
  - \log^2 \left(1 - \frac{p_3^2}{M^2}\right)
  \Bigg\}
  + O(\epsilon).
\end{align}
where we have used the results for 1-loop integrals presented in
\cite{Ellis:2007qk}. The conclusion is that $Z^\prime$ is not approximated by
$Z^\prime_n$ to order $n$, which completes the counterexample to
 eq.~\refeq{eq:tilde-redefined-sources}.

As a final comment, notice that the approximation should be recovered on-shell,
as $Z^\prime$ and $Z^\prime_n$ differ only from the original generating function
by the source terms. This can be checked directly: diagram $B$ does not have a
pole at $p_1^2 = 0$ and therefore it does not contribute to the S matrix.

\section{Gauge covariance}
\label{app:covariance}

Let $S[\phi, A]$ be a gauge-invariant action, where $A$ are the gauge
fields. We show here that its functional derivatives with respect to the fields
$\phi$ and $A$ are gauge covariant. Consider a redefinition
\begin{equation}
  \phi \to \phi' = \phi + \zeta G,
  \qquad
  A \to A' = A + \eta H,
\end{equation}
where $G$ and $H$ are covariant operators with $G$ in the same representation as
$\phi$ and $H$ in the adjoint representation. The gauge fields $A$ only appear
in the action $S$ through the field strength
$F^A_{\mu\nu} = \d_\mu A_\nu - \d_\nu A_\mu - i g [A_\mu, A_\nu]$
and the covariant derivative $D^A_\mu = \d_\mu - i g A_\mu$. For these
objects, we have
\begin{align}
  F^{A'}_{\mu \nu}
  &=
  F^A_{\mu\nu}
  + \eta \left( D^A_\mu H_\nu - D^A_\nu H_\mu \right)
  - i \eta^2 g [H_\mu, H_\nu],
  \\
  D^{A'}_\mu \Ocal
  &=
  D^A_\mu \Ocal
  - i \eta g H_\mu \Ocal.
\end{align}

All the terms in this expressions are covariant, with the same representation
under the gauge group. It follows that the transformed action
$S[\phi'(\phi, A), A'(\phi, A)]$ is still gauge
invariant. Its expansion in $\zeta$, $\eta$ is
\begin{equation}
  S[\phi + \zeta G, A + \eta H]
  =
  \sum_{m, n = 0}^\infty
  \frac{\zeta^m \, \eta^n}{(m + n)!}
  \;
  G^{\alpha_1} \cdots G^{\alpha_m}
  \;
  H^{\beta_1} \cdots H^{\beta_n}
  \;
  \frac{
    \delta^{m + n} S
  }{
    \delta \phi^{\alpha_1} \cdots \delta \phi^{\alpha_m}
    \delta A^{\beta_1} \cdots \delta A^{\beta_n}
  }.
\end{equation}

Because this is invariant for any $\zeta$ and $\eta$, it must be invariant order
by order in each of them. Now, the covariance of the functional derivatives
follows from the covariance of the product of the operators $G$ and $H$. In
particular, the equation of motion operators $\delta S/\delta \phi$ and
$\delta S/\delta A$ must be covariant and therefore it is possible to write them
in terms of field strengths and covariant derivatives, with no independent
occurrences of the gauge fields and partial derivatives.

\bibliographystyle{JHEP}
\bibliography{redefinitions}{}

\providecommand{\href}[2]{#2}\begingroup\raggedright\begin{thebibliography}{10}

\bibitem{Polchinski:1983gv}
J.~Polchinski, \emph{{Renormalization and Effective Lagrangians}},
  \href{https://doi.org/10.1016/0550-3213(84)90287-6}{\emph{Nucl. Phys.}
  {\bfseries B231} (1984) 269}.

\bibitem{Weinberg:1995mt}
S.~Weinberg, \emph{{The Quantum theory of fields. Vol. 1: Foundations}}.
  Cambridge University Press, 2005.

\bibitem{Latorre:2000qc}
J.~I. Latorre and T.~R. Morris, \emph{{Exact scheme independence}},
  \href{https://doi.org/10.1088/1126-6708/2000/11/004}{\emph{JHEP} {\bfseries
  11} (2000) 004} [\href{https://arxiv.org/abs/hep-th/0008123}{{\ttfamily
  hep-th/0008123}}].

\bibitem{Chisholm:1961tha}
J.~S.~R. Chisholm, \emph{{Change of variables in quantum field theories}},
  \href{https://doi.org/10.1016/0029-5582(61)90106-7}{\emph{Nucl. Phys.}
  {\bfseries 26} (1961) 469}.

\bibitem{Kamefuchi:1961sb}
S.~Kamefuchi, L.~O'Raifeartaigh and A.~Salam, \emph{{Change of variables and
  equivalence theorems in quantum field theories}},
  \href{https://doi.org/10.1016/0029-5582(61)90056-6,
  10.1016/0029-5582(61)91075-6}{\emph{Nucl. Phys.} {\bfseries 28} (1961) 529}.

\bibitem{Divakaran:1963yxz}
P.~P. Divakaran, \emph{{Equivalence theorems and point transformations in field
  theory}}, \href{https://doi.org/10.1016/0029-5582(63)90731-4}{\emph{Nucl.
  Phys.} {\bfseries 42} (1963) 235}.

\bibitem{Kallosh:1972ap}
R.~E. Kallosh and I.~V. Tyutin, \emph{{The Equivalence theorem and gauge
  invariance in renormalizable theories}}, {\emph{Yad. Fiz.} {\bfseries 17}
  (1973) 190}.

\bibitem{Salam:1971sp}
A.~Salam and J.~A. Strathdee, \emph{{Equivalent formulations of massive vector
  field theories}}, \href{https://doi.org/10.1103/PhysRevD.2.2869}{\emph{Phys.
  Rev.} {\bfseries D2} (1970) 2869}.

\bibitem{Ball:1993zy}
R.~D. Ball and R.~S. Thorne, \emph{{Renormalizability of effective scalar field
  theory}}, \href{https://doi.org/10.1006/aphy.1994.1109}{\emph{Annals Phys.}
  {\bfseries 236} (1994) 117}
  [\href{https://arxiv.org/abs/hep-th/9310042}{{\ttfamily hep-th/9310042}}].

\bibitem{Arzt:1993gz}
C.~Arzt, \emph{{Reduced effective Lagrangians}},
  \href{https://doi.org/10.1016/0370-2693(94)01419-D}{\emph{Phys. Lett.}
  {\bfseries B342} (1995) 189}
  [\href{https://arxiv.org/abs/hep-ph/9304230}{{\ttfamily hep-ph/9304230}}].

\bibitem{tHooft:1973wag}
G.~'t~Hooft and M.~J.~G. Veltman, \emph{{DIAGRAMMAR}}, {\emph{NATO Sci. Ser. B}
  {\bfseries 4} (1974) 177}.

\bibitem{Georgi:1991ch}
H.~Georgi, \emph{{On-shell effective field theory}},
  \href{https://doi.org/10.1016/0550-3213(91)90244-R}{\emph{Nucl. Phys.}
  {\bfseries B361} (1991) 339}.

\bibitem{Grzadkowski:2003tf}
B.~Grzadkowski, Z.~Hioki, K.~Ohkuma and J.~Wudka, \emph{{Probing anomalous top
  quark couplings induced by dimension-six operators at photon colliders}},
  \href{https://doi.org/10.1016/j.nuclphysb.2004.04.006}{\emph{Nucl. Phys.}
  {\bfseries B689} (2004) 108}
  [\href{https://arxiv.org/abs/hep-ph/0310159}{{\ttfamily hep-ph/0310159}}].

\bibitem{Fox:2007in}
P.~J. Fox, Z.~Ligeti, M.~Papucci, G.~Perez and M.~D. Schwartz,
  \emph{{Deciphering top flavor violation at the LHC with $B$ factories}},
  \href{https://doi.org/10.1103/PhysRevD.78.054008}{\emph{Phys. Rev.}
  {\bfseries D78} (2008) 054008}
  [\href{https://arxiv.org/abs/0704.1482}{{\ttfamily 0704.1482}}].

\bibitem{AguilarSaavedra:2008zc}
J.~A. Aguilar-Saavedra, \emph{{A Minimal set of top anomalous couplings}},
  \href{https://doi.org/10.1016/j.nuclphysb.2008.12.012}{\emph{Nucl. Phys.}
  {\bfseries B812} (2009) 181}
  [\href{https://arxiv.org/abs/0811.3842}{{\ttfamily 0811.3842}}].

\bibitem{AguilarSaavedra:2009mx}
J.~A. Aguilar-Saavedra, \emph{{A Minimal set of top-Higgs anomalous
  couplings}},
  \href{https://doi.org/10.1016/j.nuclphysb.2009.06.022}{\emph{Nucl. Phys.}
  {\bfseries B821} (2009) 215}
  [\href{https://arxiv.org/abs/0904.2387}{{\ttfamily 0904.2387}}].

\bibitem{Grzadkowski:2010es}
B.~Grzadkowski, M.~Iskrzynski, M.~Misiak and J.~Rosiek, \emph{{Dimension-Six
  Terms in the Standard Model Lagrangian}},
  \href{https://doi.org/10.1007/JHEP10(2010)085}{\emph{JHEP} {\bfseries 10}
  (2010) 085} [\href{https://arxiv.org/abs/1008.4884}{{\ttfamily 1008.4884}}].

\bibitem{Alfaro:1989rx}
J.~Alfaro and P.~H. Damgaard, \emph{{Field Transformations, Collective
  Coordinates and {BRST} Invariance}},
  \href{https://doi.org/10.1016/0003-4916(90)90230-L}{\emph{Annals Phys.}
  {\bfseries 202} (1990) 398}.

\bibitem{Manohar:1997qy}
A.~V. Manohar, \emph{{The HQET / NRQCD Lagrangian to order alpha / $m^3$}},
  \href{https://doi.org/10.1103/PhysRevD.56.230}{\emph{Phys. Rev.} {\bfseries
  D56} (1997) 230} [\href{https://arxiv.org/abs/hep-ph/9701294}{{\ttfamily
  hep-ph/9701294}}].

\bibitem{Jenkins:2017dyc}
E.~E. Jenkins, A.~V. Manohar and P.~Stoffer, \emph{{Low-Energy Effective Field
  Theory below the Electroweak Scale: Anomalous Dimensions}},
  \href{https://doi.org/10.1007/JHEP01(2018)084}{\emph{JHEP} {\bfseries 01}
  (2018) 084} [\href{https://arxiv.org/abs/1711.05270}{{\ttfamily
  1711.05270}}].

\bibitem{Barzinji:2018xvu}
A.~Barzinji, M.~Trott and A.~Vasudevan, \emph{{Equations of Motion for the
  Standard Model Effective Field Theory: Theory and Applications}},
  \href{https://arxiv.org/abs/1806.06354v2}{{\ttfamily 1806.06354v2}}.

\bibitem{Arzt:1994gp}
C.~Arzt, M.~B. Einhorn and J.~Wudka, \emph{{Patterns of deviation from the
  standard model}},
  \href{https://doi.org/10.1016/0550-3213(94)00336-D}{\emph{Nucl. Phys.}
  {\bfseries B433} (1995) 41}
  [\href{https://arxiv.org/abs/hep-ph/9405214}{{\ttfamily hep-ph/9405214}}].

\bibitem{Hays:2018zze}
C.~Hays, A.~Martin, V.~Sanz and J.~Setford, \emph{{On the impact of
  dimension-eight SMEFT operators on Higgs measurements}},
  \href{https://arxiv.org/abs/1808.00442}{{\ttfamily 1808.00442}}.

\bibitem{Passarino:2016saj}
G.~Passarino, \emph{{Field reparametrization in effective field theories}},
  \href{https://doi.org/10.1140/epjp/i2017-11291-5}{\emph{Eur. Phys. J. Plus}
  {\bfseries 132} (2017) 16}
  [\href{https://arxiv.org/abs/1610.09618}{{\ttfamily 1610.09618}}].

\bibitem{Gervais:1976ws}
J.-L. Gervais and A.~Jevicki, \emph{{Point Canonical Transformations in Path
  Integral}}, \href{https://doi.org/10.1016/0550-3213(76)90422-3}{\emph{Nucl.
  Phys.} {\bfseries B110} (1976) 93}.

\bibitem{Breitenlohner:1977hr}
P.~Breitenlohner and D.~Maison, \emph{{Dimensional Renormalization and the
  Action Principle}}, \href{https://doi.org/10.1007/BF01609069}{\emph{Commun.
  Math. Phys.} {\bfseries 52} (1977) 11}.

\bibitem{Lehmann:1954rq}
H.~Lehmann, K.~Symanzik and W.~Zimmermann, \emph{{On the formulation of
  quantized field theories}},
  \href{https://doi.org/10.1007/BF02731765}{\emph{Nuovo Cim.} {\bfseries 1}
  (1955) 205}.

\bibitem{Vilkovisky:1984st}
G.~A. Vilkovisky, \emph{{The Unique Effective Action in Quantum Field Theory}},
  \href{https://doi.org/10.1016/0550-3213(84)90228-1}{\emph{Nucl. Phys.}
  {\bfseries B234} (1984) 125}.

\bibitem{Anselmi:2012jt}
D.~Anselmi, \emph{{A Master Functional For Quantum Field Theory}},
  \href{https://doi.org/10.1140/epjc/s10052-013-2385-y}{\emph{Eur. Phys. J.}
  {\bfseries C73} (2013) 2385}
  [\href{https://arxiv.org/abs/1205.3584}{{\ttfamily 1205.3584}}].

\bibitem{Denner:2014zga}
A.~Denner and J.-N. Lang, \emph{{The Complex-Mass Scheme and Unitarity in
  perturbative Quantum Field Theory}},
  \href{https://doi.org/10.1140/epjc/s10052-015-3579-2}{\emph{Eur. Phys. J.}
  {\bfseries C75} (2015) 377}
  [\href{https://arxiv.org/abs/1406.6280}{{\ttfamily 1406.6280}}].

\bibitem{Anselmi:2012aq}
D.~Anselmi, \emph{{A General Field-Covariant Formulation Of Quantum Field
  Theory}}, \href{https://doi.org/10.1140/epjc/s10052-013-2338-5}{\emph{Eur.
  Phys. J.} {\bfseries C73} (2013) 2338}
  [\href{https://arxiv.org/abs/1205.3279}{{\ttfamily 1205.3279}}].

\bibitem{Shore:1990wp}
G.~M. Shore, \emph{{New methods for the renormalization of composite operator
  Green functions}},
  \href{https://doi.org/10.1016/0550-3213(91)90557-E}{\emph{Nucl. Phys.}
  {\bfseries B362} (1991) 85}.

\bibitem{Lizana:2017sjz}
J.~M. Lizana and M.~P\'erez-Victoria, \emph{{Wilsonian renormalisation of CFT
  correlation functions: Field theory}},
  \href{https://doi.org/10.1007/JHEP06(2017)139}{\emph{JHEP} {\bfseries 06}
  (2017) 139} [\href{https://arxiv.org/abs/1702.07773}{{\ttfamily
  1702.07773}}].

\bibitem{Bonneau:1985ea}
G.~Bonneau and F.~Delduc, \emph{{Nonlinear Renormalization and the Equivalence
  Theorem}}, \href{https://doi.org/10.1016/0550-3213(86)90184-7}{\emph{Nucl.
  Phys.} {\bfseries B266} (1986) 536}.

\bibitem{Politzer:1980me}
H.~D. Politzer, \emph{{Power Corrections at Short Distances}},
  \href{https://doi.org/10.1016/0550-3213(80)90172-8}{\emph{Nucl. Phys.}
  {\bfseries B172} (1980) 349}.

\bibitem{KlubergStern:1975hc}
H.~Kluberg-Stern and J.~B. Zuber, \emph{{Renormalization of Nonabelian Gauge
  Theories in a Background Field Gauge. 2. Gauge Invariant Operators}},
  \href{https://doi.org/10.1103/PhysRevD.12.3159}{\emph{Phys. Rev.} {\bfseries
  D12} (1975) 3159}.

\bibitem{GrosseKnetter:1993td}
C.~Grosse-Knetter, \emph{{Effective Lagrangians with higher derivatives and
  equations of motion}},
  \href{https://doi.org/10.1103/PhysRevD.49.6709}{\emph{Phys. Rev.} {\bfseries
  D49} (1994) 6709} [\href{https://arxiv.org/abs/hep-ph/9306321}{{\ttfamily
  hep-ph/9306321}}].

\bibitem{Wudka:1994ny}
J.~Wudka, \emph{{Electroweak effective Lagrangians}},
  \href{https://doi.org/10.1142/S0217751X94000959}{\emph{Int. J. Mod. Phys.}
  {\bfseries A9} (1994) 2301}
  [\href{https://arxiv.org/abs/hep-ph/9406205}{{\ttfamily hep-ph/9406205}}].

\bibitem{Manohar:2018aog}
A.~V. Manohar, \emph{{Introduction to Effective Field Theories}},  in
  \emph{{Les Houches summer school: EFT in Particle Physics and Cosmology Les
  Houches, Chamonix Valley, France, July 3-28, 2017}}, 2018,
  \href{https://arxiv.org/abs/1804.05863}{{\ttfamily 1804.05863}}.

\bibitem{Anselmi:2006yh}
D.~Anselmi, \emph{{Renormalization and causality violations in classical
  gravity coupled with quantum matter}},
  \href{https://doi.org/10.1088/1126-6708/2007/01/062}{\emph{JHEP} {\bfseries
  01} (2007) 062} [\href{https://arxiv.org/abs/hep-th/0605205}{{\ttfamily
  hep-th/0605205}}].

\bibitem{delAguila:2016zcb}
F.~del Aguila, Z.~Kunszt and J.~Santiago, \emph{{One-loop effective lagrangians
  after matching}},
  \href{https://doi.org/10.1140/epjc/s10052-016-4081-1}{\emph{Eur. Phys. J.}
  {\bfseries C76} (2016) 244}
  [\href{https://arxiv.org/abs/1602.00126}{{\ttfamily 1602.00126}}].

\bibitem{Beneke:1997zp}
M.~Beneke and V.~A. Smirnov, \emph{{Asymptotic expansion of Feynman integrals
  near threshold}},
  \href{https://doi.org/10.1016/S0550-3213(98)00138-2}{\emph{Nucl. Phys.}
  {\bfseries B522} (1998) 321}
  [\href{https://arxiv.org/abs/hep-ph/9711391}{{\ttfamily hep-ph/9711391}}].

\bibitem{Fuentes-Martin:2016uol}
J.~Fuentes-Martin, J.~Portoles and P.~Ruiz-Femenia, \emph{{Integrating out
  heavy particles with functional methods: a simplified framework}},
  \href{https://doi.org/10.1007/JHEP09(2016)156}{\emph{JHEP} {\bfseries 09}
  (2016) 156} [\href{https://arxiv.org/abs/1607.02142}{{\ttfamily
  1607.02142}}].

\bibitem{Bilenky:1993bt}
M.~S. Bilenky and A.~Santamaria, \emph{{One loop effective Lagrangian for a
  standard model with a heavy charged scalar singlet}},
  \href{https://doi.org/10.1016/0550-3213(94)90375-1}{\emph{Nucl. Phys.}
  {\bfseries B420} (1994) 47}
  [\href{https://arxiv.org/abs/hep-ph/9310302}{{\ttfamily hep-ph/9310302}}].

\bibitem{Bilenky:1994kt}
M.~S. Bilenky and A.~Santamaria, \emph{{Beyond the standard model with
  effective lagrangians}},  in \emph{{28th International Symposium on Particle
  Theory Wendisch-Rietz, Germany, August 30-September 3, 1994}}, pp.~215--224,
  1994, \href{https://arxiv.org/abs/hep-ph/9503257}{{\ttfamily
  hep-ph/9503257}}.

\bibitem{Henning:2016lyp}
B.~Henning, X.~Lu and H.~Murayama, \emph{{One-loop Matching and Running with
  Covariant Derivative Expansion}},
  \href{https://doi.org/10.1007/JHEP01(2018)123}{\emph{JHEP} {\bfseries 01}
  (2018) 123} [\href{https://arxiv.org/abs/1604.01019}{{\ttfamily
  1604.01019}}].

\bibitem{Meetz:1969as}
K.~Meetz, \emph{{Realization of chiral symmetry in a curved isospin space}},
  \href{https://doi.org/10.1063/1.1664881}{\emph{J. Math. Phys.} {\bfseries 10}
  (1969) 589}.

\bibitem{Honerkamp:1996va}
J.~Honerkamp and K.~Meetz, \emph{{Chiral-invariant perturbation theory}},
  \href{https://doi.org/10.1103/PhysRevD.3.1996}{\emph{Phys. Rev.} {\bfseries
  D3} (1971) 1996}.

\bibitem{Honerkamp:1971sh}
J.~Honerkamp, \emph{{Chiral multiloops}},
  \href{https://doi.org/10.1016/0550-3213(72)90299-4}{\emph{Nucl. Phys.}
  {\bfseries B36} (1972) 130}.

\bibitem{Ecker:1972bm}
G.~Ecker and J.~Honerkamp, \emph{{Application of invariant renormalization to
  the nonlinear chiral invariant pion lagrangian in the one-loop
  approximation}},
  \href{https://doi.org/10.1016/0550-3213(71)90468-8}{\emph{Nucl. Phys.}
  {\bfseries B35} (1971) 481}.

\bibitem{AlvarezGaume:1981hn}
L.~Alvarez-Gaume, D.~Z. Freedman and S.~Mukhi, \emph{{The Background Field
  Method and the Ultraviolet Structure of the Supersymmetric Nonlinear Sigma
  Model}}, \href{https://doi.org/10.1016/0003-4916(81)90006-3}{\emph{Annals
  Phys.} {\bfseries 134} (1981) 85}.

\bibitem{AlvarezGaume:1981hm}
L.~Alvarez-Gaume and D.~Z. Freedman, \emph{{Geometrical Structure and
  Ultraviolet Finiteness in the Supersymmetric Sigma Model}},
  \href{https://doi.org/10.1007/BF01208280}{\emph{Commun. Math. Phys.}
  {\bfseries 80} (1981) 443}.

\bibitem{Boulware:1981ns}
D.~G. Boulware and L.~S. Brown, \emph{{SYMMETRIC SPACE SCALAR FIELD THEORY}},
  \href{https://doi.org/10.1016/0003-4916(82)90192-0}{\emph{Annals Phys.}
  {\bfseries 138} (1982) 392}.

\bibitem{Alonso:2016oah}
R.~Alonso, E.~E. Jenkins and A.~V. Manohar, \emph{{Geometry of the Scalar
  Sector}}, \href{https://doi.org/10.1007/JHEP08(2016)101}{\emph{JHEP}
  {\bfseries 08} (2016) 101}
  [\href{https://arxiv.org/abs/1605.03602}{{\ttfamily 1605.03602}}].

\bibitem{Coleman:1969sm}
S.~R. Coleman, J.~Wess and B.~Zumino, \emph{{Structure of phenomenological
  Lagrangians. 1.}},
  \href{https://doi.org/10.1103/PhysRev.177.2239}{\emph{Phys. Rev.} {\bfseries
  177} (1969) 2239}.

\bibitem{Manohar:1983md}
A.~Manohar and H.~Georgi, \emph{{Chiral Quarks and the Nonrelativistic Quark
  Model}}, \href{https://doi.org/10.1016/0550-3213(84)90231-1}{\emph{Nucl.
  Phys.} {\bfseries B234} (1984) 189}.

\bibitem{Jenkins:2013sda}
E.~E. Jenkins, A.~V. Manohar and M.~Trott, \emph{{Naive Dimensional Analysis
  Counting of Gauge Theory Amplitudes and Anomalous Dimensions}},
  \href{https://doi.org/10.1016/j.physletb.2013.09.020}{\emph{Phys. Lett.}
  {\bfseries B726} (2013) 697}
  [\href{https://arxiv.org/abs/1309.0819}{{\ttfamily 1309.0819}}].

\bibitem{Gavela:2016bzc}
B.~M. Gavela, E.~E. Jenkins, A.~V. Manohar and L.~Merlo, \emph{{Analysis of
  General Power Counting Rules in Effective Field Theory}},
  \href{https://doi.org/10.1140/epjc/s10052-016-4332-1}{\emph{Eur. Phys. J.}
  {\bfseries C76} (2016) 485}
  [\href{https://arxiv.org/abs/1601.07551}{{\ttfamily 1601.07551}}].

\bibitem{Giudice:2007fh}
G.~F. Giudice, C.~Grojean, A.~Pomarol and R.~Rattazzi, \emph{{The
  Strongly-Interacting Light Higgs}},
  \href{https://doi.org/10.1088/1126-6708/2007/06/045}{\emph{JHEP} {\bfseries
  06} (2007) 045} [\href{https://arxiv.org/abs/hep-ph/0703164}{{\ttfamily
  hep-ph/0703164}}].

\bibitem{Brivio:2017vri}
I.~Brivio and M.~Trott, \emph{{The Standard Model as an Effective Field
  Theory}},  \href{https://arxiv.org/abs/1706.08945}{{\ttfamily 1706.08945}}.

\bibitem{deBlas:2017xtg}
J.~de~Blas, J.~C. Criado, M.~P\'erez-Victoria and J.~Santiago, \emph{{Effective
  description of general extensions of the Standard Model: the complete
  tree-level dictionary}},
  \href{https://doi.org/10.1007/JHEP03(2018)109}{\emph{JHEP} {\bfseries 03}
  (2018) 109} [\href{https://arxiv.org/abs/1711.10391}{{\ttfamily
  1711.10391}}].

\bibitem{Helset:2017mlf}
A.~Helset and M.~Trott, \emph{{On interference and non-interference in the
  SMEFT}}, \href{https://doi.org/10.1007/JHEP04(2018)038}{\emph{JHEP}
  {\bfseries 04} (2018) 038}
  [\href{https://arxiv.org/abs/1711.07954}{{\ttfamily 1711.07954}}].

\bibitem{AguilarSaavedra:2010sq}
J.~A. Aguilar-Saavedra, \emph{{Effective operators in top physics}},
  \href{https://doi.org/10.22323/1.120.0378}{\emph{PoS} {\bfseries ICHEP2010}
  (2010) 378} [\href{https://arxiv.org/abs/1008.3225}{{\ttfamily 1008.3225}}].

\bibitem{Azatov:2016sqh}
A.~Azatov, R.~Contino, C.~S. Machado and F.~Riva, \emph{{Helicity selection
  rules and noninterference for BSM amplitudes}},
  \href{https://doi.org/10.1103/PhysRevD.95.065014}{\emph{Phys. Rev.}
  {\bfseries D95} (2017) 065014}
  [\href{https://arxiv.org/abs/1607.05236}{{\ttfamily 1607.05236}}].

\bibitem{AguilarSaavedra:2011vw}
J.~A. Aguilar-Saavedra and M.~P\'erez-Victoria, \emph{{Probing the Tevatron t
  tbar asymmetry at LHC}},
  \href{https://doi.org/10.1007/JHEP05(2011)034}{\emph{JHEP} {\bfseries 05}
  (2011) 034} [\href{https://arxiv.org/abs/1103.2765}{{\ttfamily 1103.2765}}].

\bibitem{Maldacena:1997re}
J.~M. Maldacena, \emph{{The Large N limit of superconformal field theories and
  supergravity}}, \href{https://doi.org/10.1023/A:1026654312961,
  10.4310/ATMP.1998.v2.n2.a1}{\emph{Int. J. Theor. Phys.} {\bfseries 38} (1999)
  1113} [\href{https://arxiv.org/abs/hep-th/9711200}{{\ttfamily
  hep-th/9711200}}].

\bibitem{Collins:1984xc}
J.~C. Collins, \emph{{Renormalization}}, vol.~26 of \emph{Cambridge Monographs
  on Mathematical Physics}. Cambridge University Press, Cambridge, 1986,
  \href{https://doi.org/10.1017/CBO9780511622656}{10.1017/CBO9780511622656}.

\bibitem{Einhorn:2001kj}
M.~B. Einhorn and J.~Wudka, \emph{{Effective beta functions for effective field
  theory}}, \href{https://doi.org/10.1088/1126-6708/2001/08/025}{\emph{JHEP}
  {\bfseries 08} (2001) 025}
  [\href{https://arxiv.org/abs/hep-ph/0105035}{{\ttfamily hep-ph/0105035}}].

\bibitem{Grojean:2013kd}
C.~Grojean, E.~E. Jenkins, A.~V. Manohar and M.~Trott, \emph{{Renormalization
  Group Scaling of Higgs Operators and $\Gamma(h -> \gamma \gamma)$}},
  \href{https://doi.org/10.1007/JHEP04(2013)016}{\emph{JHEP} {\bfseries 04}
  (2013) 016} [\href{https://arxiv.org/abs/1301.2588}{{\ttfamily 1301.2588}}].

\bibitem{Elias-Miro:2013gya}
J.~Elias-Mir\'o, J.~R. Espinosa, E.~Masso and A.~Pomarol,
  \emph{{Renormalization of dimension-six operators relevant for the Higgs
  decays $h\rightarrow \gamma\gamma,\gamma Z$}},
  \href{https://doi.org/10.1007/JHEP08(2013)033}{\emph{JHEP} {\bfseries 08}
  (2013) 033} [\href{https://arxiv.org/abs/1302.5661}{{\ttfamily 1302.5661}}].

\bibitem{Elias-Miro:2013mua}
J.~Elias-Mir\'o, J.~R. Espinosa, E.~Masso and A.~Pomarol, \emph{{Higgs windows
  to new physics through d=6 operators: constraints and one-loop anomalous
  dimensions}}, \href{https://doi.org/10.1007/JHEP11(2013)066}{\emph{JHEP}
  {\bfseries 11} (2013) 066} [\href{https://arxiv.org/abs/1308.1879}{{\ttfamily
  1308.1879}}].

\bibitem{Jenkins:2013zja}
E.~E. Jenkins, A.~V. Manohar and M.~Trott, \emph{{Renormalization Group
  Evolution of the Standard Model Dimension Six Operators I: Formalism and
  lambda Dependence}},
  \href{https://doi.org/10.1007/JHEP10(2013)087}{\emph{JHEP} {\bfseries 10}
  (2013) 087} [\href{https://arxiv.org/abs/1308.2627}{{\ttfamily 1308.2627}}].

\bibitem{Jenkins:2013wua}
E.~E. Jenkins, A.~V. Manohar and M.~Trott, \emph{{Renormalization Group
  Evolution of the Standard Model Dimension Six Operators II: Yukawa
  Dependence}}, \href{https://doi.org/10.1007/JHEP01(2014)035}{\emph{JHEP}
  {\bfseries 01} (2014) 035} [\href{https://arxiv.org/abs/1310.4838}{{\ttfamily
  1310.4838}}].

\bibitem{Elias-Miro:2013eta}
J.~Elias-Mir\'o, C.~Grojean, R.~S. Gupta and D.~Marzocca, \emph{{Scaling and
  tuning of EW and Higgs observables}},
  \href{https://doi.org/10.1007/JHEP05(2014)019}{\emph{JHEP} {\bfseries 05}
  (2014) 019} [\href{https://arxiv.org/abs/1312.2928}{{\ttfamily 1312.2928}}].

\bibitem{Alonso:2014rga}
R.~Alonso, E.~E. Jenkins and A.~V. Manohar, \emph{{Holomorphy without
  Supersymmetry in the Standard Model Effective Field Theory}},
  \href{https://doi.org/10.1016/j.physletb.2014.10.045}{\emph{Phys. Lett.}
  {\bfseries B739} (2014) 95}
  [\href{https://arxiv.org/abs/1409.0868}{{\ttfamily 1409.0868}}].

\bibitem{Elias-Miro:2014eia}
J.~Elias-Mir\'o, J.~R. Espinosa and A.~Pomarol, \emph{{One-loop
  non-renormalization results in EFTs}},
  \href{https://doi.org/10.1016/j.physletb.2015.05.056}{\emph{Phys. Lett.}
  {\bfseries B747} (2015) 272}
  [\href{https://arxiv.org/abs/1412.7151}{{\ttfamily 1412.7151}}].

\bibitem{deBlas:2012qp}
J.~de~Blas, J.~M. Lizana and M.~P\'erez-Victoria, \emph{{Combining searches of
  Z' and W' bosons}},
  \href{https://doi.org/10.1007/JHEP01(2013)166}{\emph{JHEP} {\bfseries 01}
  (2013) 166} [\href{https://arxiv.org/abs/1211.2229}{{\ttfamily 1211.2229}}].

\bibitem{Pappadopulo:2014qza}
D.~Pappadopulo, A.~Thamm, R.~Torre and A.~Wulzer, \emph{{Heavy Vector Triplets:
  Bridging Theory and Data}},
  \href{https://doi.org/10.1007/JHEP09(2014)060}{\emph{JHEP} {\bfseries 09}
  (2014) 060} [\href{https://arxiv.org/abs/1402.4431}{{\ttfamily 1402.4431}}].

\bibitem{Aaboud:2018bun}
{\scshape ATLAS} collaboration, M.~Aaboud et~al., \emph{{Combination of
  searches for heavy resonances decaying into bosonic and leptonic final states
  using 36 fb$^{-1}$ of proton-proton collision data at $\sqrt{s} = 13$ TeV
  with the ATLAS detector}},
  \href{https://doi.org/10.1103/PhysRevD.98.052008}{\emph{Phys. Rev.}
  {\bfseries D98} (2018) 052008}
  [\href{https://arxiv.org/abs/1808.02380}{{\ttfamily 1808.02380}}].

\bibitem{Criado:2017khh}
J.~C. Criado, \emph{{MatchingTools: a Python library for symbolic effective
  field theory calculations}},
  \href{https://doi.org/10.1016/j.cpc.2018.02.016}{\emph{Comput. Phys. Commun.}
  {\bfseries 227} (2018) 42}
  [\href{https://arxiv.org/abs/1710.06445}{{\ttfamily 1710.06445}}].

\bibitem{Bakshi:2018ics}
S.~Das~Bakshi, J.~Chakrabortty and S.~K. Patra, \emph{{CoDEx: Wilson
  coefficient calculator connecting SMEFT to UV theory}},
  \href{https://arxiv.org/abs/1808.04403}{{\ttfamily 1808.04403}}.

\bibitem{Falkowski:2015wza}
A.~Falkowski, B.~Fuks, K.~Mawatari, K.~Mimasu, F.~Riva and V.~Sanz,
  \emph{{Rosetta: an operator basis translator for Standard Model effective
  field theory}},
  \href{https://doi.org/10.1140/epjc/s10052-015-3806-x}{\emph{Eur. Phys. J.}
  {\bfseries C75} (2015) 583}
  [\href{https://arxiv.org/abs/1508.05895}{{\ttfamily 1508.05895}}].

\bibitem{Gripaios:2018zrz}
B.~Gripaios and D.~Sutherland, \emph{{DEFT: A program for operators in EFT}},
  \href{https://arxiv.org/abs/1807.07546}{{\ttfamily 1807.07546}}.

\bibitem{Ellis:2007qk}
R.~K. Ellis and G.~Zanderighi, \emph{{Scalar one-loop integrals for QCD}},
  \href{https://doi.org/10.1088/1126-6708/2008/02/002}{\emph{JHEP} {\bfseries
  02} (2008) 002} [\href{https://arxiv.org/abs/0712.1851}{{\ttfamily
  0712.1851}}].

\end{thebibliography}\endgroup

\end{document}